\pdfoutput=1
\documentclass[fleqn,usenatbib]{mnras}

\usepackage{newtxtext,newtxmath}

\usepackage[T1]{fontenc}

\DeclareRobustCommand{\VAN}[3]{#2}
\let\VANthebibliography\thebibliography
\def\thebibliography{\DeclareRobustCommand{\VAN}[3]{##3}\VANthebibliography}

\usepackage{graphicx}	
\usepackage{amsmath}	
\usepackage[normalem]{ulem}

\title[Data driven photo-z from SNeIa light curves]{Data driven photometric redshift estimation from Type Ia Supernovae light curves}

\author[F. M. F. de Oliveira et al.]{
Felipe M. F. de Oliveira$^{1},$
Marcelo Vargas dos Santos$^{2},$
Ribamar R. R. Reis$^{3,4}$\\
$^{1}$ Hurb Technologies, 22776-090, Rio de Janeiro, RJ, Brazil \\
$^{2}$ Unidade Acadêmica de Física, Universidade Federal de Campina Grande, 58429-900 Campina Grande, PB, Brazil \\
$^{3}$ Instituto de F\'isica, Universidade Federal do Rio de Janeiro, 21941-972, Rio de Janeiro, RJ, Brazil \\
$^{4}$ Observat\'orio do Valongo, Universidade Federal do Rio de Janeiro, 20080-090, Rio de Janeiro, RJ, Brazil
}

\date{Accepted XXX. Received YYY; in original form ZZZ}

\pubyear{2022}

\begin{document}
\label{firstpage}
\pagerange{\pageref{firstpage}--\pageref{lastpage}}
\maketitle

\begin{abstract}
Redshift measurement has always been a constant need in modern Astronomy and Cosmology. And as new surveys have been providing an immense amount of data on astronomical objects, the need to process such data automatically proves to be increasingly necessary. In this article we use simulated data from the Dark Energy Survey (DES) and from a pipeline originally created to classify supernovae, we developed a linear regression algorithm optimized through novel automated machine learning (AutoML) frameworks achieving an error score better than ordinary data preprocessing methods when compared with other modern algorithms (such as XGBoost). Numerically, the photometric prediction RMSE of Ia supernovae was reduced from 0.16 to 0.09 and the RMSE of all supernovae types decreased from 0.20 to 0.14.
Our pipeline consists of 4 steps: through spectroscopic data points we interpolate the light curve using gaussian process fitting algorithm, then using a wavelet transform we extract the most important features of such curves, in sequence we reduce the dimensionality of such features through principal component analysis (PCA), and in the end we applied super learning techniques (stacked ensemble methods) through an AutoML framework dedicated to optimize the parameters of several different machine learning models, better resolving the problem.
As a final check, we obtained probability distribution functions (PDFs) using gaussian kernel density estimations (kdes) through the predictions of more than 50 models trained and optimized by AutoML. Those PDFs were calculated to replicate the original curves which used SALT2 model, a model used for the simulation of the raw data itself.
\end{abstract}

\begin{keywords}
techniques: photometric -- transients: supernovae -- cosmology: miscellaneous -- cosmology: observations -- software: data analysis
\end{keywords}



\section{Introduction}

As Cosmology has entered an era of massive data and information \citep{Ishida2019MachineLA}, the need for more elaborated machine learning techniques has been increasing due to the fields that needed its application, such as Multifield Cosmology \citep{villaescusa2021multifield}, Light curve classifications \citep{burhanudin2021light} and Big Data challenges over astronomical surveys \citep{hlovzek2020results}. 

In the supernova field, most of the efforts have been focused on the problem of photometric classification \citep{dobryakov2021photometric}, \citep{leoni2021fink}, since only type Ia events are established as a suitable cosmological probe. Current and future projects aim to detect thousands of supernovae and sample contamination due to failed classification is a major source of systematics for these experiments.
The basic approach for these classification algorithms usually include the use of redshift, measured or estimated, from the host galaxy \citep{leoni2021fink}. However, this may not be feasible for high redshift events, in which the host can be too faint to be detected.

In this work we are interested in photometric redshift estimation from the supernova observations themselves \citep{mitra2021cosmology}, taking benefit from the framework already built for the classification problem. We seek to investigate how classic machine learning pre-processing techniques can be combined with one of the most recent automated machine learning (AutoML) frameworks \citep{He_2021} to enable the choice of the best algorithms, in order to significantly minimize the prediction error. We also used the final predictions alongside with SALT2 \citep{Guy:2007dv} model to "re-generate" the curves and assess our model comparing the newly generated data with the original simulated raw data.

This study's baseline was an already existing pipeline from \citep{Lochner:2016hbn}, in which the same data was pre-processed in order to generate features for a supervised classification algorithm. Through these existing features, we focus on another variable of the same data set, the redshift. From these pre-processed features developed by the feature engineering \citep{brownlee2014} from the previous pipeline, we proceeded to a supervised linear regression problem, whose target variable was the redshift. The idea of using this baseline was to take advantage of the previous study developed, so it was only necessary to focus on the optimization of linear regression models for redshift prediction.

In order to decide which models, parameters and hyperparameters better suit our problem, we used an AutoML framework named H2O \citep{H2OAutoML20}. AutoML frameworks have been recently developed in order to optimize and automatize models and hyperparemeters selections. They consist in empirical strategies and benchmarked techniques which facilitate the convergence for optimal hyperparameters and models. With H2O AutoML framework, it was possible to obtain metric evaluations' results for the $n$ best models from different families of algorithms \autoref{fig:amlboard50}. In this article we set $n$ value to 50. 

Additionally, as we needed a probability distribution function to retrace the SALT2 curve, we developed other 2 models capable of offering not only a single value of regression, but also a probability distribution function. Those two models were developed from a super learning stacked ensemble \citep{Naimi172395} technique, where it selected the best model for each type of input to predict the redshift value of that sample.
The first model considered the entire pool of 50 models and the second one considered only the best models of each family. With those 2 stacked ensemble models we were able to offer not only the best prediction value for each input from our pool of models, but also offer
a probability distribution function of the result.

In this paper we contextualize the data and expose the results of the scores and the models' board from the AutoML framework. In \autoref{sec:spcc} we summarize the data set used. In \autoref{sec:mas} we explain how raw data was disposed and all the pipeline steps. And finally we present our results in \autoref{sec:res} and conclusions in \autoref{sec:con}.

\section{Photometric data}\label{sec:spcc}

In this work we use simulated data from the Supernova Photometric Classification Challenge (SNPCC), proposed by~\citep{Kessler:2010wk} in order to stimulate the development of tools to address this problem. The data include a mix of simulated SNe, with the different types selected in proportion to their expected rate and as if they had been measured with the \textit{griz} filters of the DES with realistic observing conditions (sky noise, point spread function and atmospheric transparency) based on years of recorded conditions at the DES site. Simulations of non-Ia SNe are based on spectroscopically confirmed light curves that include non-Ia samples donated by the Carnegie Supernova Project (CSP), the Supernova Legacy Survey (SNLS), and the Sloan Digital Sky Survey-II (SDSS–II).

The SNPCC catalog contains 21319 supernova light curves corresponding to five seasons of observations, simulated by the \textsc{SNANA} software. This set is composed by 5086 SNeIa and 16231 core collapse SNe, distributed among 7 different types: II, IIn, IIp, IIL, Ib, Ib/c and Ic. The type Ia supernovae were simulated using an equal mix of the models MLCS2k2~\citep{Jha:2006fm} and SALT2~\citep{Guy:2007dv} with an additional random color variation. An extinction correction, MLCS-U2 was used in order to make the models agree in the ultraviolet~\citep{Kessler:2009ys}. The non-Ia supernovae light curves were based on spectroscopically confirmed observed ones and constructed by warping a standard spectral template to match the photometry. The SN rates were based on the results from \citep{Dilday:2008wp} and \citep{Bazin:2009}. Ten groups addressed the challenge using different algorithms including template matching and ML techniques. The results of the challenge were published in \citep{Kessler:2010qj}: the template matching code PSNID from~\citep{Sako:2007ms} was found to be the best one based on the figure-of-merit employed (which was a function of purity and completeness).

After the end of the SNPCC the catalog was unblinded, and it now serves as a very useful tool to train and test ML algorithms relying on supervised training, in which a subset of the data has known classification \emph{a priori}. In our case in particular this corresponds to assuming we will have a subset of the SNe observed spectroscopically.

Although there are more recent simulated data, we chose to use SNPCC in order to benefit from the results obtained in the work of \citep{Santos_2020} on supernova classification. In this sense, the present paper can be considered as a natural extension to the previous one.

\section{The machine learning Approach}\label{sec:mas}

The goal of a machine learning regression model is to be able to predict a certain value from predetermined features. It receives a data set with the features and a specific target variable to be predicted. The model trains over this data set and becomes able to predict the target variable value for a new set of these features.

In this article we consider the spectroscopy redshift as the target value and the features as the 20 PCA results from the pipeline of \citep{Santos_2020} article. Briefly, the pipeline consisted on interpolating the photometric light flux points with a gaussian process fitting method, next the interpolated curves are decomposed by a wavelets method, and finally, to avoid working with a huge number of features, a principal component analysis (PCA) \citep{doi:10.1080/14786440109462720} is applied to reduce data dimensionality before heading to machine learning methods.

\subsection{AutoML}

Machine learning is already known as a powerful tool for predictions and classifications when dealing with massive data sets. However, due to the increasing number of machine learning techniques \citep{hastie2009} involving neural networks, forest like algorithms, linear regressions (Lasso, Ridge, Elastic Net), XGBoost variants, suport vector machines and many other besides, it is difficult to test and to decide which technique suits better for each specific problem. To solve this issue, automated machine learning (AutoML) techniques were used through H2O library \citep{h2oSuperLearner} to choose the best model and the best parameters. Summarily, "the field of automated machine learning (AutoML) aims to develop methods that build suitable machine learning models without (or as little as possible) human intervention" \citep{Tuggener_2019}.

Some benefits that AutoML can provide are: faster automatic response, feedback on the quality of data, good baselines for almost any kind of problem and no preference for templates (no bias).

Usually, AutoML frameworks do their hyperparameter optimization and model selection through grid searches (\autoref{sub:grid}).

\subsection{H2O Framework}

H2O was chosen due to the fact that it is an open-source distributed machine learning platform designed to perform well and fast regardless large volumes of data and it is supported for other languages. So, in the future it would be easy to reproduce the code under different conditions, like alternative languages (R, Java and Scala) or utilization of graphics processing units (GPU).

H2O has some pre process techniques implemented within it. By default it treats missing data and normalizes the data set. Then it works in two phases:
\begin{enumerate}
    \item \textbf{Base Models}: First, it trains a base of pre-selected models from different families, called Base Models. In this step, H2O also searches for different sets of hyperparameters among the base models by \textbf{grid searching} it.
    
    \item \textbf{Stacked Ensembles}: Secondly, it creates two Stacked Ensembles (or super learners) \citep{Naimi172395} from the base models. One considering the set of all base models, and another one considering the set of the best models from each family.
\end{enumerate}

\subsubsection{Grid Search}\label{sub:grid}

H2O grid search starts from a series of models and hyperparameters pre selected inside its search space. Those models and hyperparameters, as well as their ranges, were chosen after an extensive benchmark process \citep{H2OAutoML20}. 

H2O supports two types of grid search – traditional (or “cartesian”) grid search and random grid search.
The cartesian one \citep{liashchynskyi2019grid} receives the set of values for each hyperparameter then it trains a model for every combination of the hyperparameter values.
In random grid search \citep{JMLR:v13:bergstra12a}, it receives the hyperparameter space in the exact same way, except H2O will sample uniformly from the set of all possible hyperparameter value combinations obeying a stopping criterion.

Once the grid search is complete, the models can be sorted by any specified particular performance metric (like RMSE or MAE). All models are stored in the H2O local server and are accessible by model id.

The default models and their hyperparameters, set by H2O benchmark process, are:

\begin{enumerate}
    \item 3 pre-specified XGBoost GBM (Grandient Boosting Machine) Models
    \item Fixed grid of H2O GLMs (General Linear Model)
    \item H2O standard Random Forest (DRF)
    \item 5 pre-pecified GBMs
    \item A near-standard Deep Neural Net and a random grid of H2o Deep Neural Nets
    \item H2O Extremely Randomized Trees (XRT) model
    \item Random grid of XGBoost GBMs
    \item Random grid of H2O GBMs
\end{enumerate}

The models are trained in a predefined order and strategy to optimize time. That order was set to test with models that consistently get good results, pre-specified XGBoost models. After XGBoost, it tests the GLM Fixed Grid \citep{osawa2011bagging}. Then, AutoML focuses on increasing the diversity of models for better performance of the ensembles, training a fixed set of Random Forests, GBMs \citep{chen2015xgboost} and Deep Neural Networks (Deep Learning). Finally, it performs a random search to get new models.

\subsubsection{Stacked Ensembles}

\begin{figure}
\includegraphics[width=0.9\columnwidth]{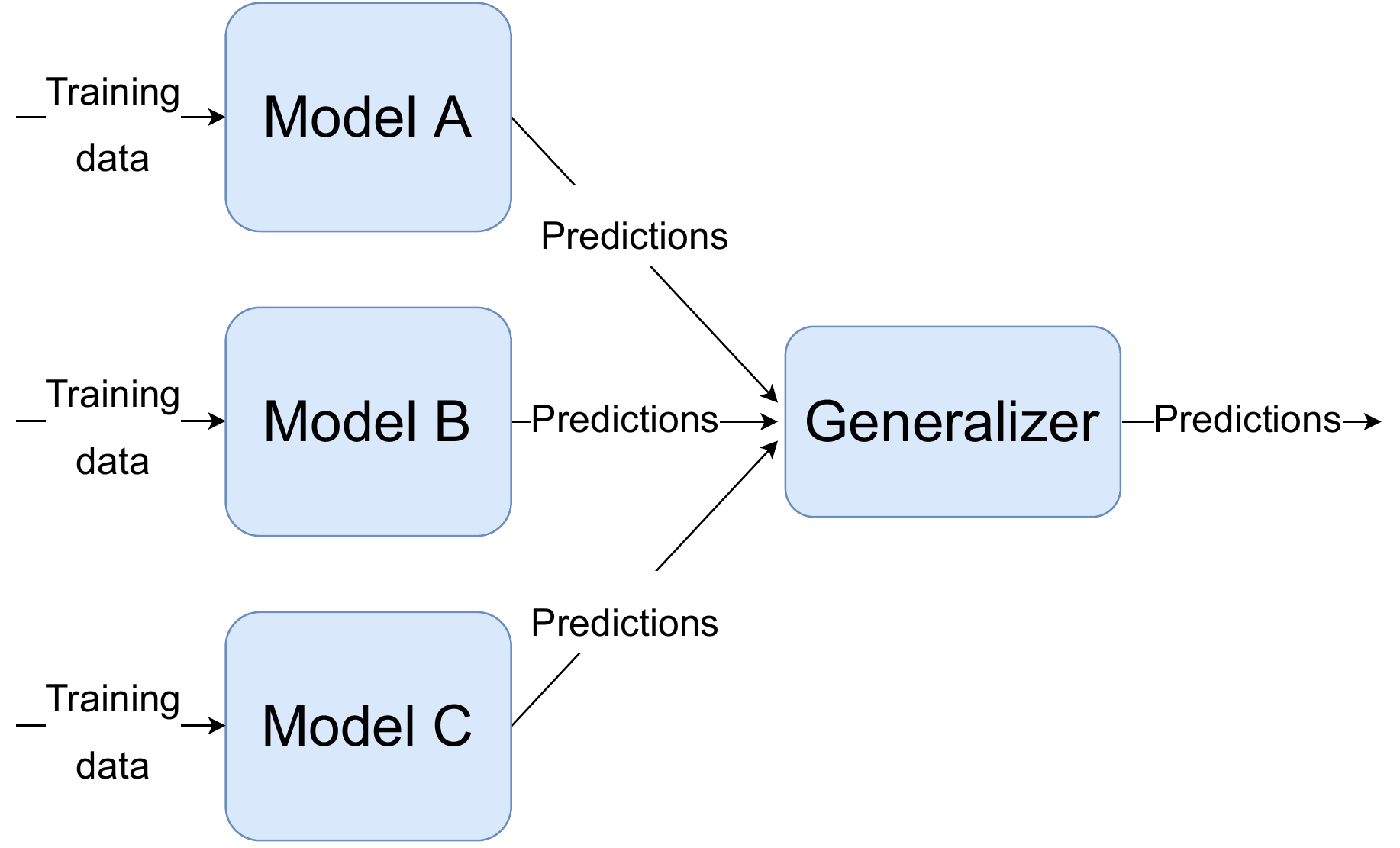}
\caption{Diagram illustrating a general ensemble generalizer.}
\label{fig:ensemble}
\end{figure}

Briefly, ensembles (\autoref{fig:ensemble}) are machine learning methods that use multiple models to obtain a better prediction than it would be obtained by each one of the models individually.

Stacking (or super learners \citep{vanderLaanPolleyHubbard+2007}) is a class of algorithms that involves training one model in addition. It is called metalearner. Its objective is to find the optimal combination of the base models. A metalearner (or combiner) is an algorithm trained over the data set with predictions of base models' cross-validation.

The following steps represent the tasks involving training a super learner. All those tasks are automated inside H2O \citep{h2oSuperLearner}.

\begin{enumerate}
    \item Set up the ensemble.
        \begin{enumerate}
            \item Specify a list of L base algorithms (with a specific set of model parameters).
            \item Specify a metalearning algorithm.
        \end{enumerate}

    \item Train the ensemble.
        \begin{enumerate}
            \item Train each of the L base algorithms on the training set.
            \item Perform k-fold cross-validation on each of these learners and collect the cross-validated predicted values from each of the L algorithms.
            \item The N cross-validated predicted values from each of the L algorithms can be combined to form a new N x L matrix. This matrix, along with the original response vector, is called the “level-one” data. (N = number of rows in the training set.)
            \item Train the metalearning algorithm on the level-one data. The “ensemble model” consists of the L base learning models and the metalearning model, which can then be used to generate predictions on a test set.
        \end{enumerate}

\end{enumerate}

\begin{figure}
\includegraphics[width=0.9\columnwidth]{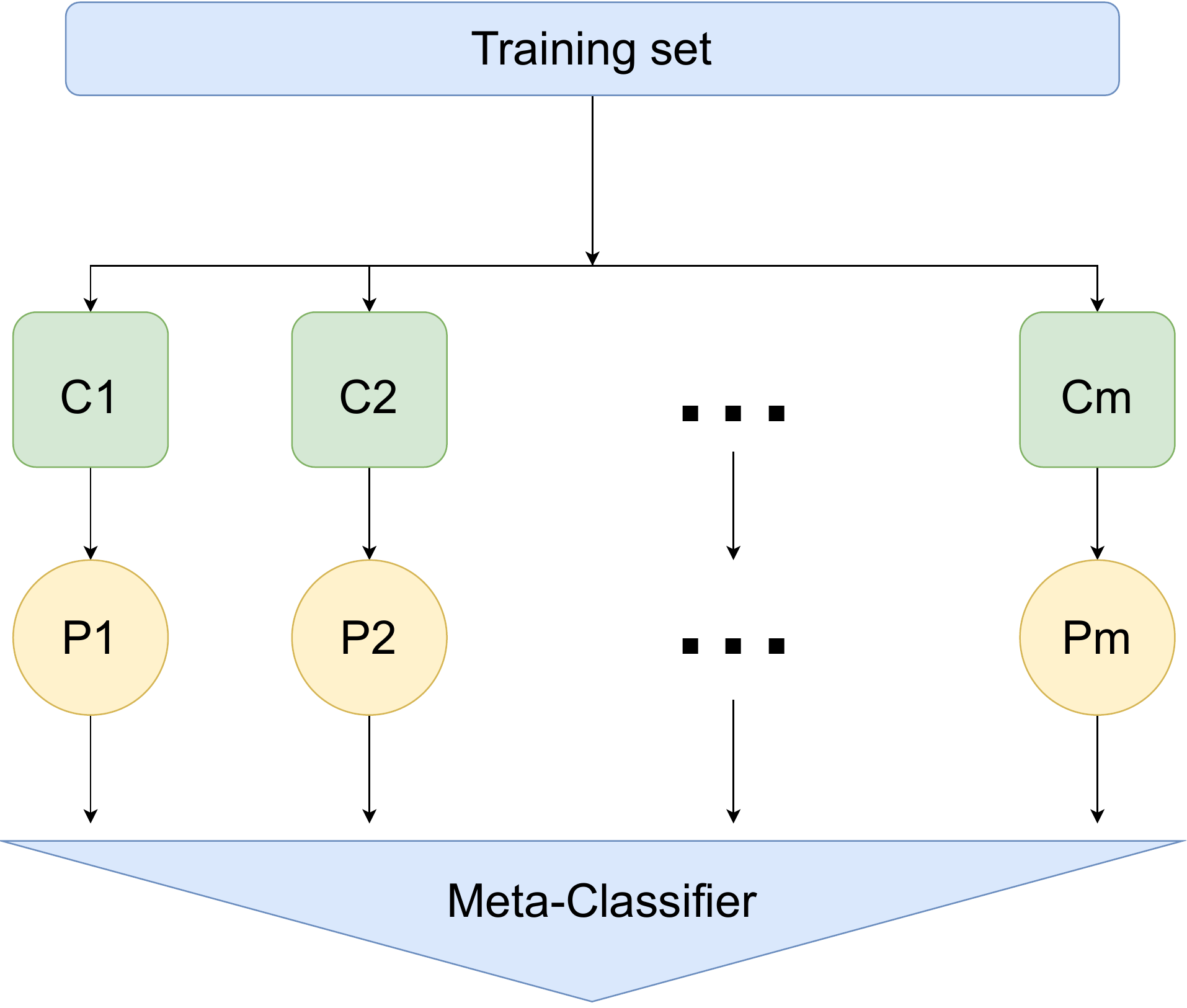}
\caption{Super learning diagram representing all the strong learners predicting their values and the meta-classifier choosing the final one.}
\label{fig:superlearning}
\end{figure}

A diagram of a classification model is represented in \autoref{fig:superlearning}.

\subsection{Modeling AutoML}

Before using H2O to model our regression, we divided our data set in two sub sets to be trained and tested. The first one considered only type Ia supernovae and the other one all objects from SNPCC.

After applying all the pipeline from \citep{Santos_2020} and \citep{Lochner:2016hbn} (Gaussian process fitting, wavelets transform and PCA) we inserted our 20 most significant principal components in H2O AutoML train, splitting the inputted data set into two sets, the validation frame and the training frame. The rate we followed was 0.5, so approximately half of our objects were used on training/test cross-validation and the other half was used for evaluation and measuring the metrics. \footnote{Note that  the training frame will be divided following a K-Fold cross validation ratio of 0.8, dividing into 5 folds and crossing it, obtaining the leaderboards in \autoref{app:leaderboards}, the validation frame is an entire unknown data for the model, its purpose is to evaluate the model's performance on unknown data, and it will be analysed in \autoref{sec:res}.}

After testing 50 different models, we obtained 2 stacked ensemble, one with the best models of each family and other one with all 50 models regardless the family and parameters. The best one was the one considering 50 models.

Once we had a 50 models ensemble it was possible to obtain a gaussian KDE probability density function from it, a necessary information to fit the light curve following SALT2 model.

\subsection{Predictors Gaussian KDE}

\begin{figure*}
    \centering
    \includegraphics[width=\columnwidth]{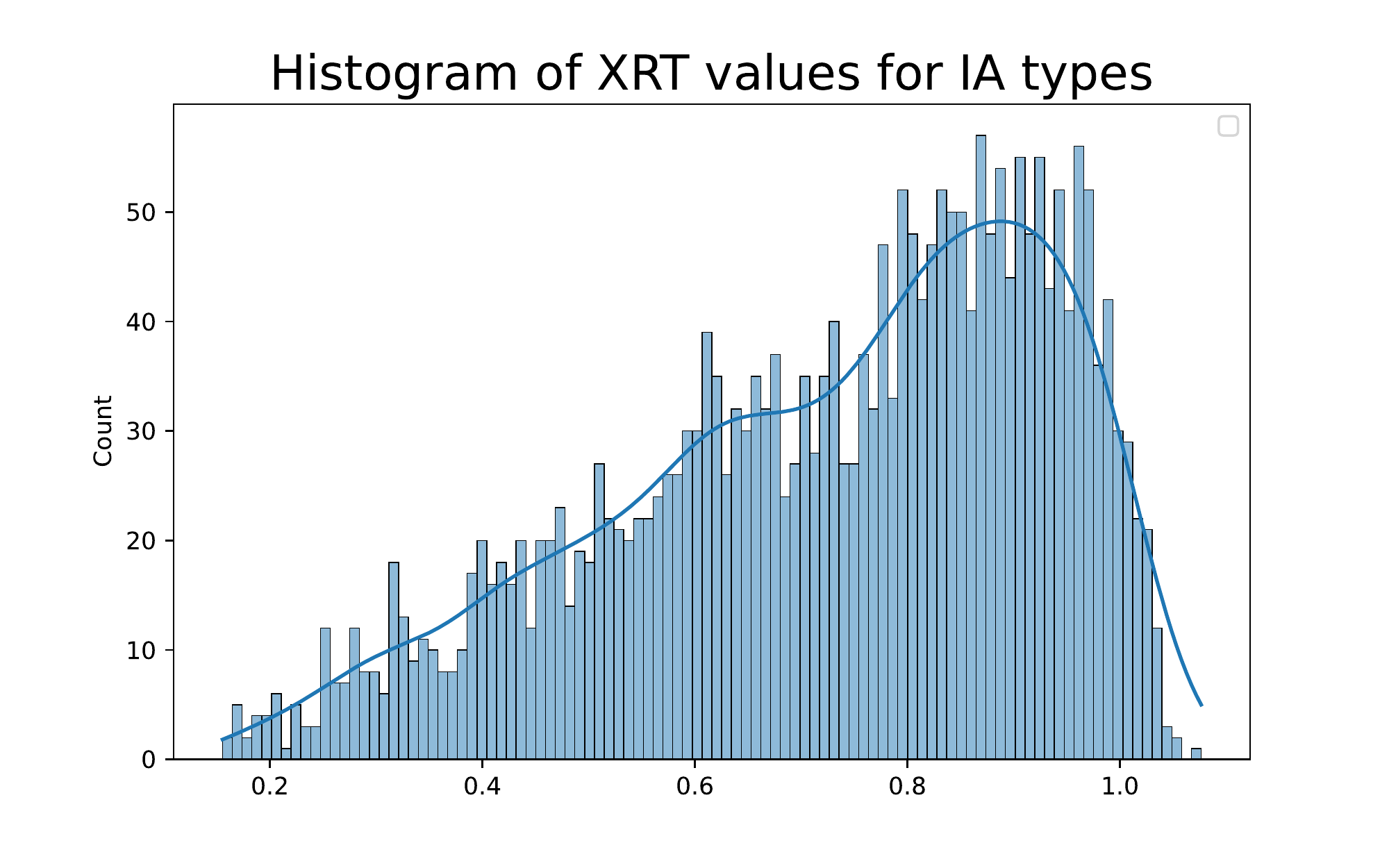}
    \includegraphics[width=\columnwidth]{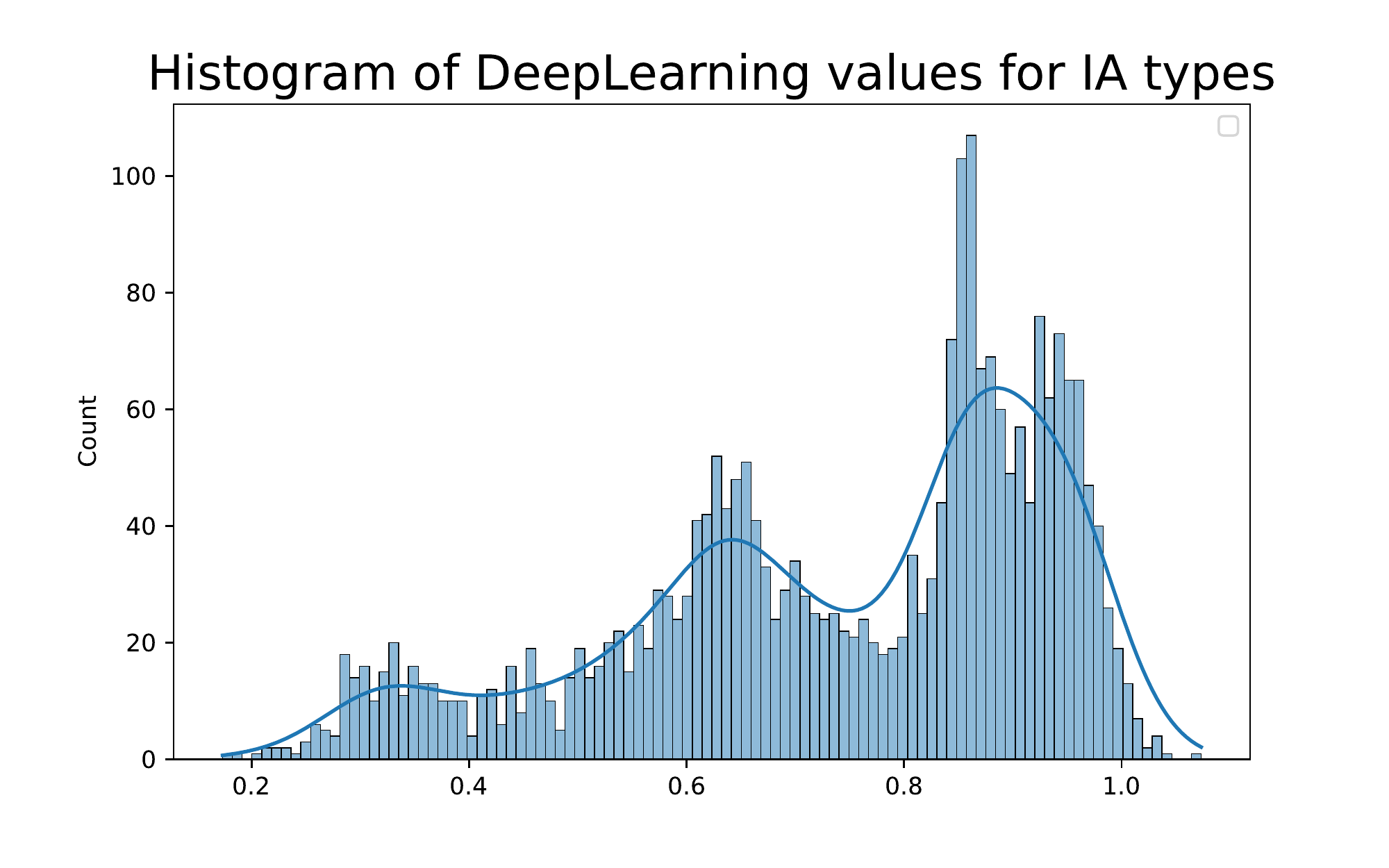}\\
    \includegraphics[width=\columnwidth]{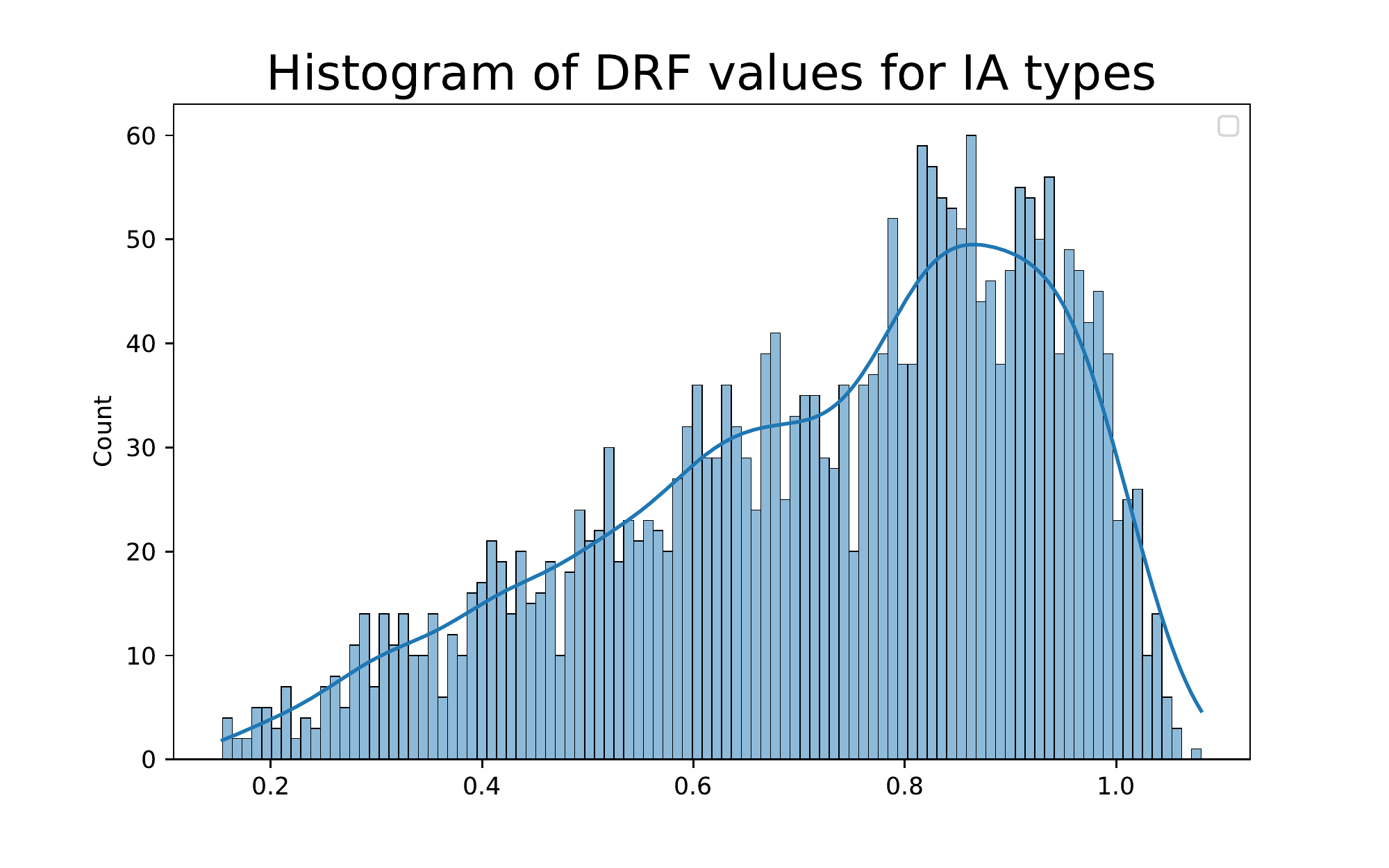}
    \includegraphics[width=\columnwidth]{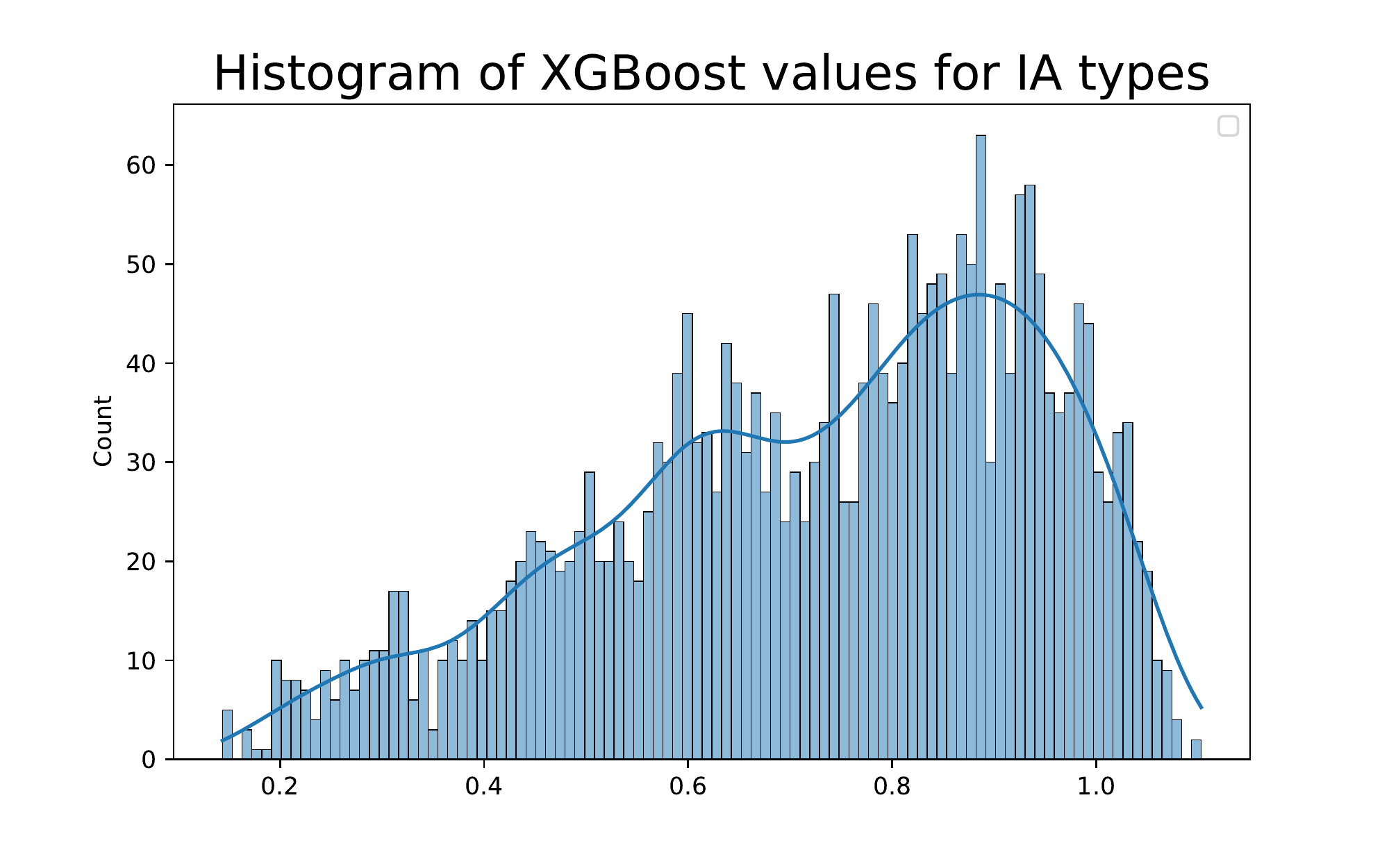}\\
    \includegraphics[width=\columnwidth]{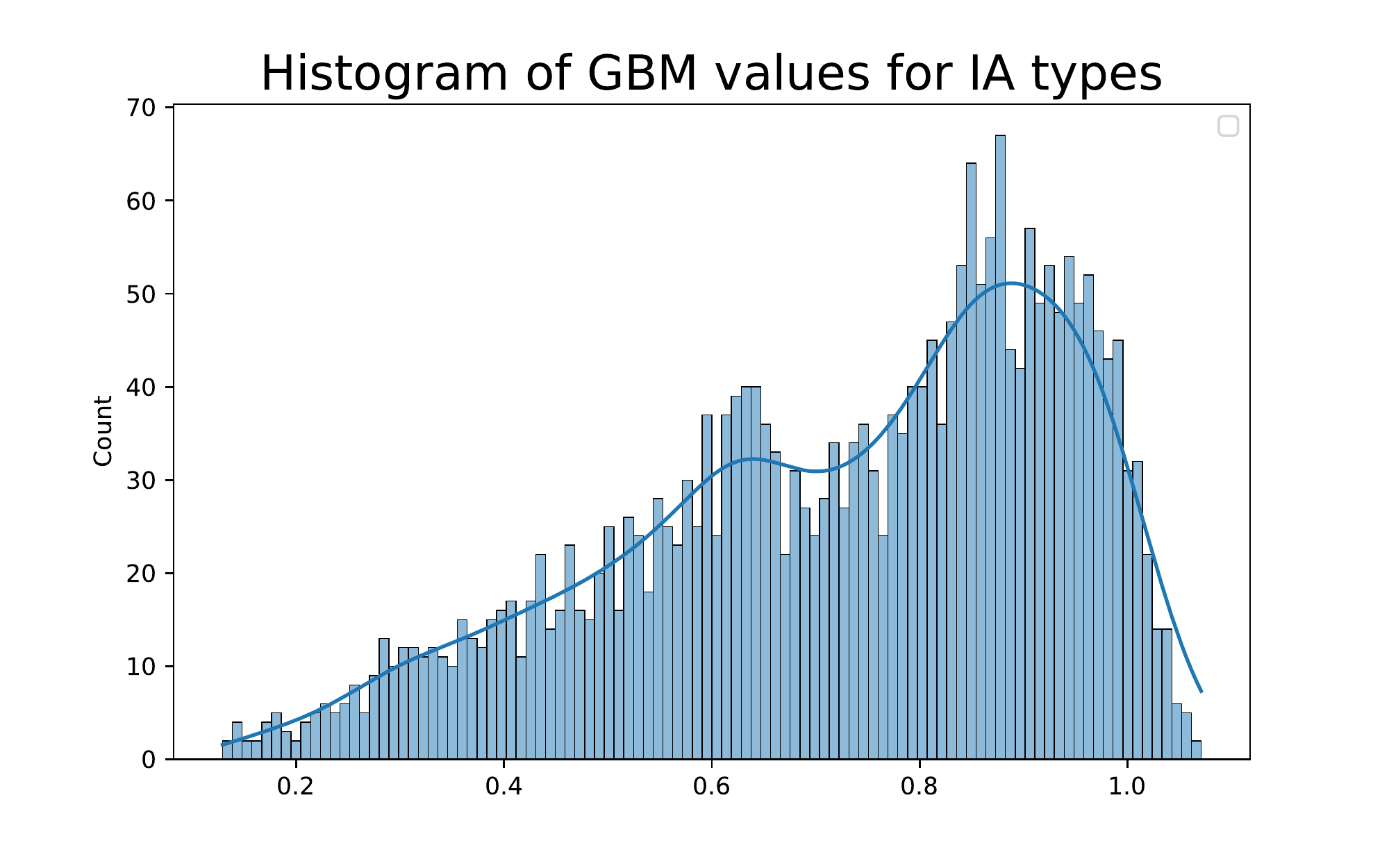}
    \includegraphics[width=\columnwidth]{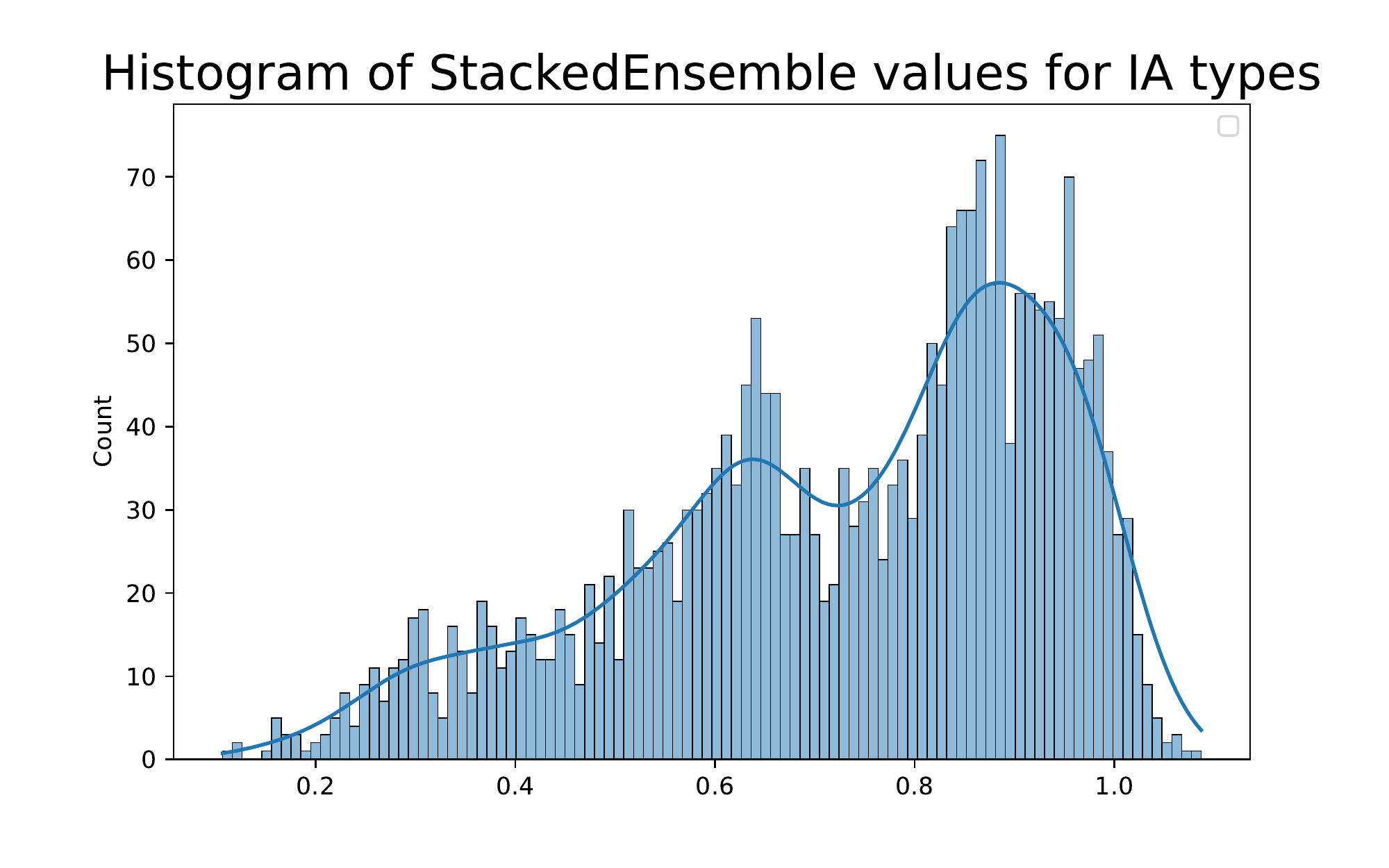}\\
    \includegraphics[width=\columnwidth]{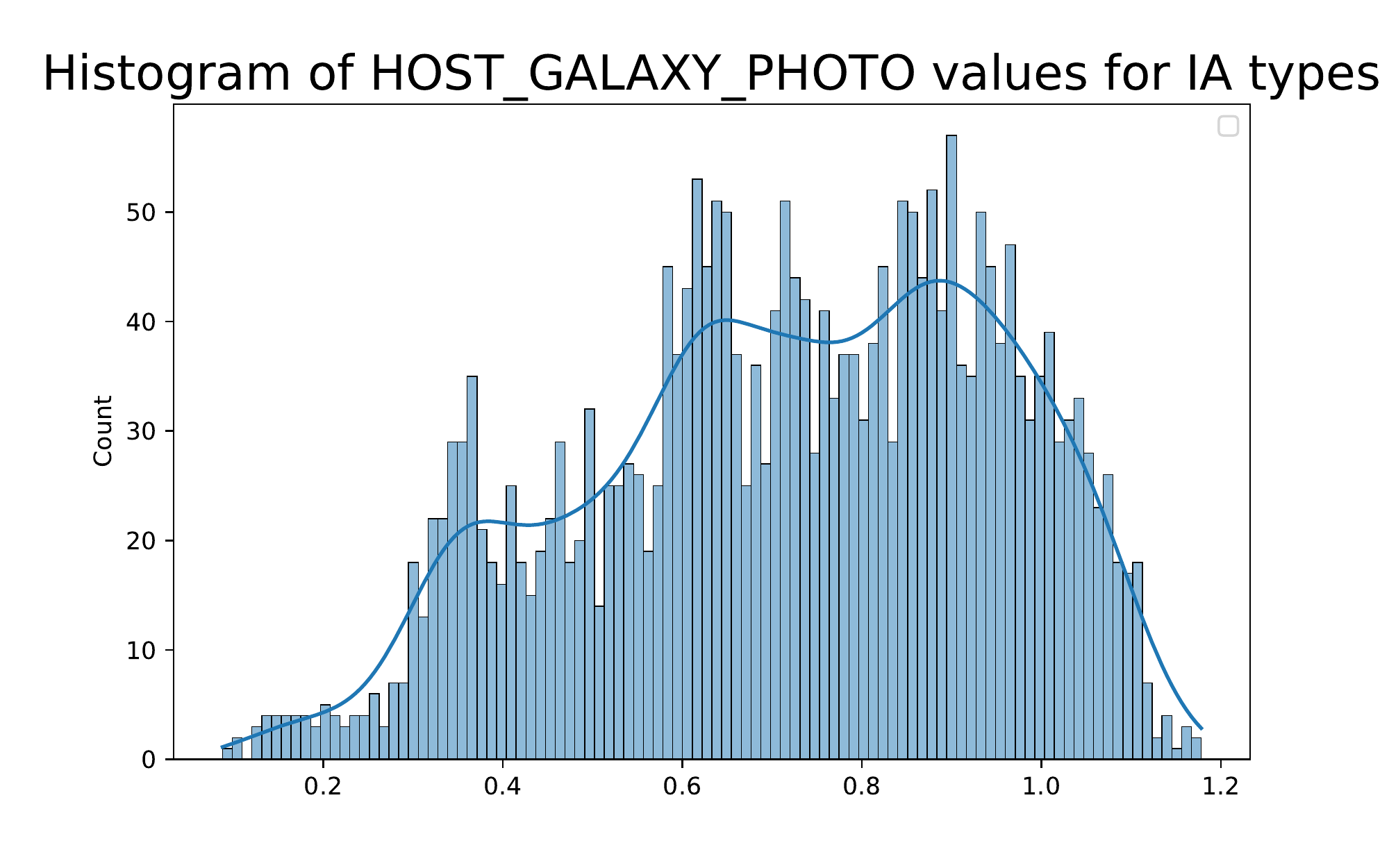}
    \includegraphics[width=\columnwidth]{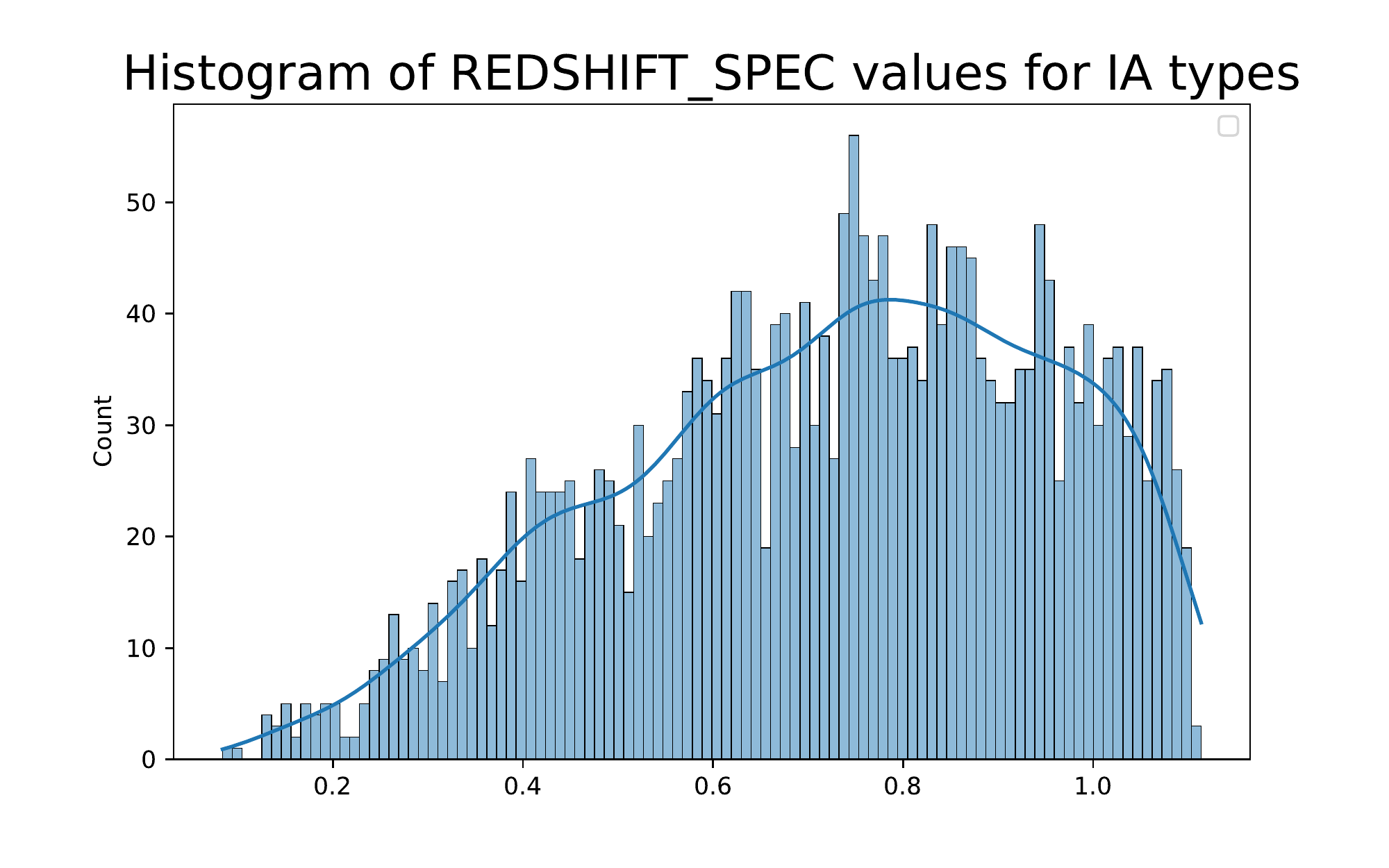}\\
    \caption{Histogram of redshift predictions values for best model from each family (Top 6 figures), and histograms with photometric and spectroscopic redshift (last 2 figures).}
    \label{fig:histograms}
\end{figure*}

Once we got the 52 models (50 default "grid searched" models plus 2 stacked ensembles), we predict the redshift value of each object. The results of the predictions from the best model of each family are illustrated in \autoref{fig:histograms} as a histogram of redshift values.

\begin{figure*}
    \centering
    \includegraphics[width=\columnwidth]{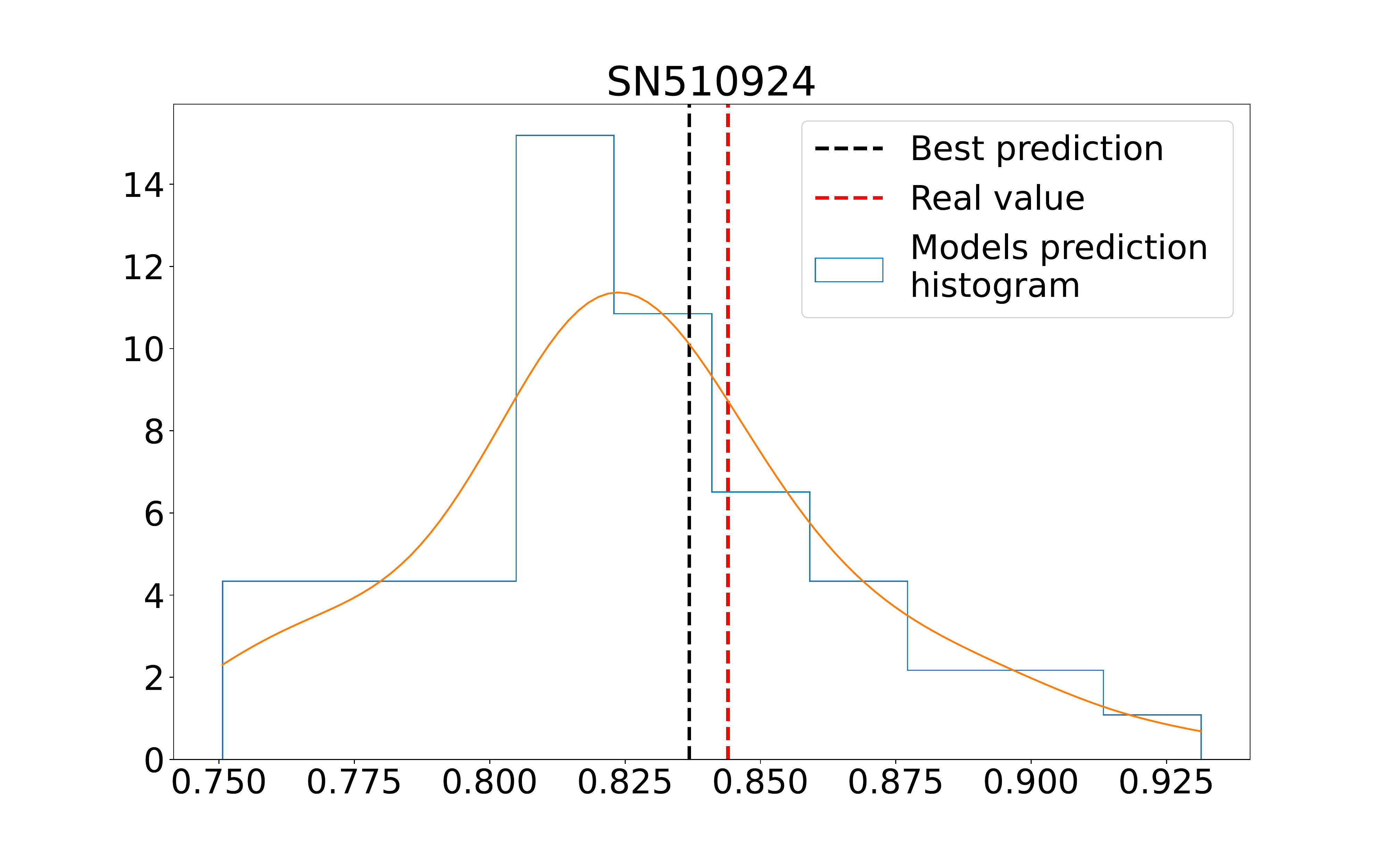}
    \includegraphics[width=\columnwidth]{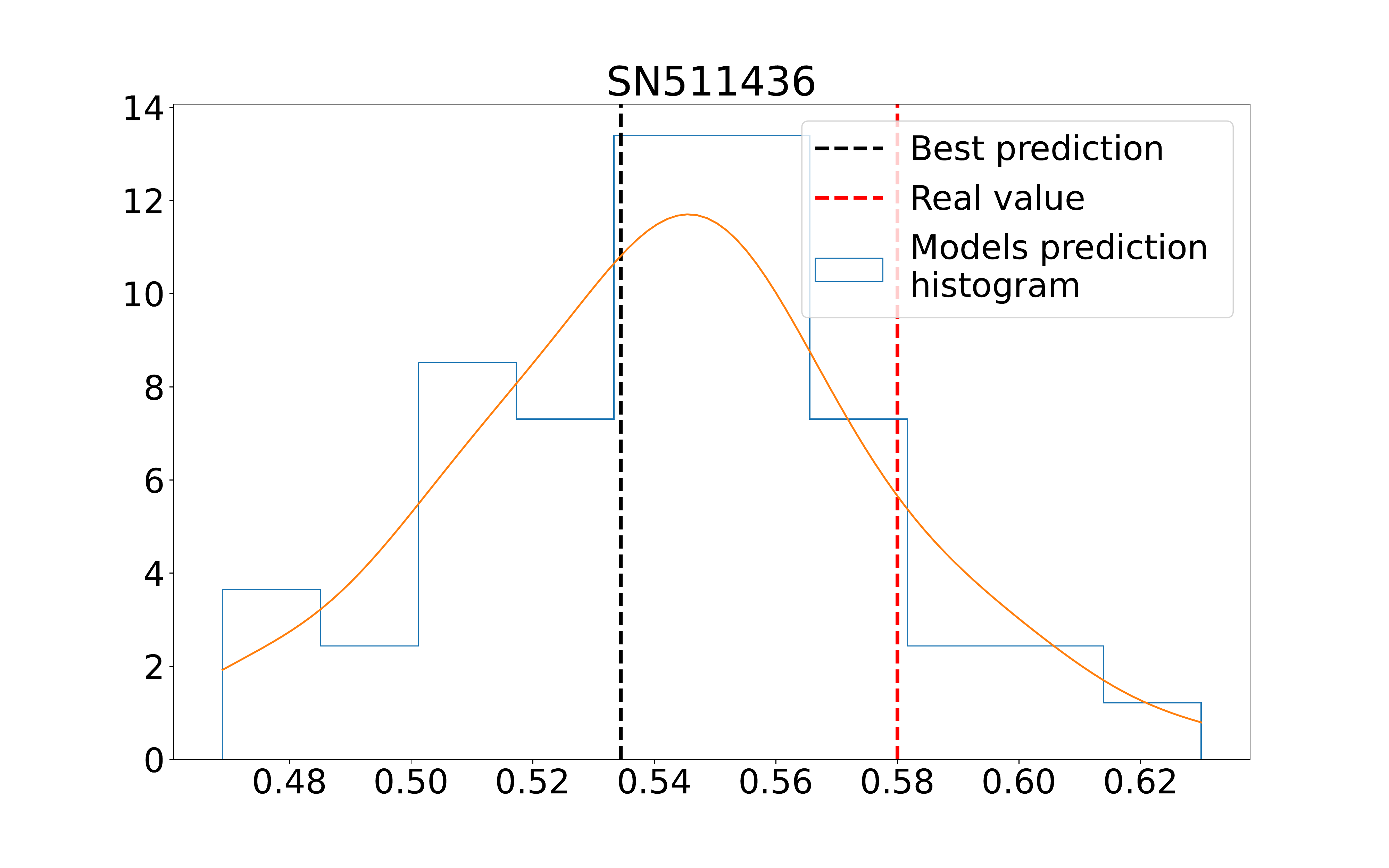}\\
    \includegraphics[width=\columnwidth]{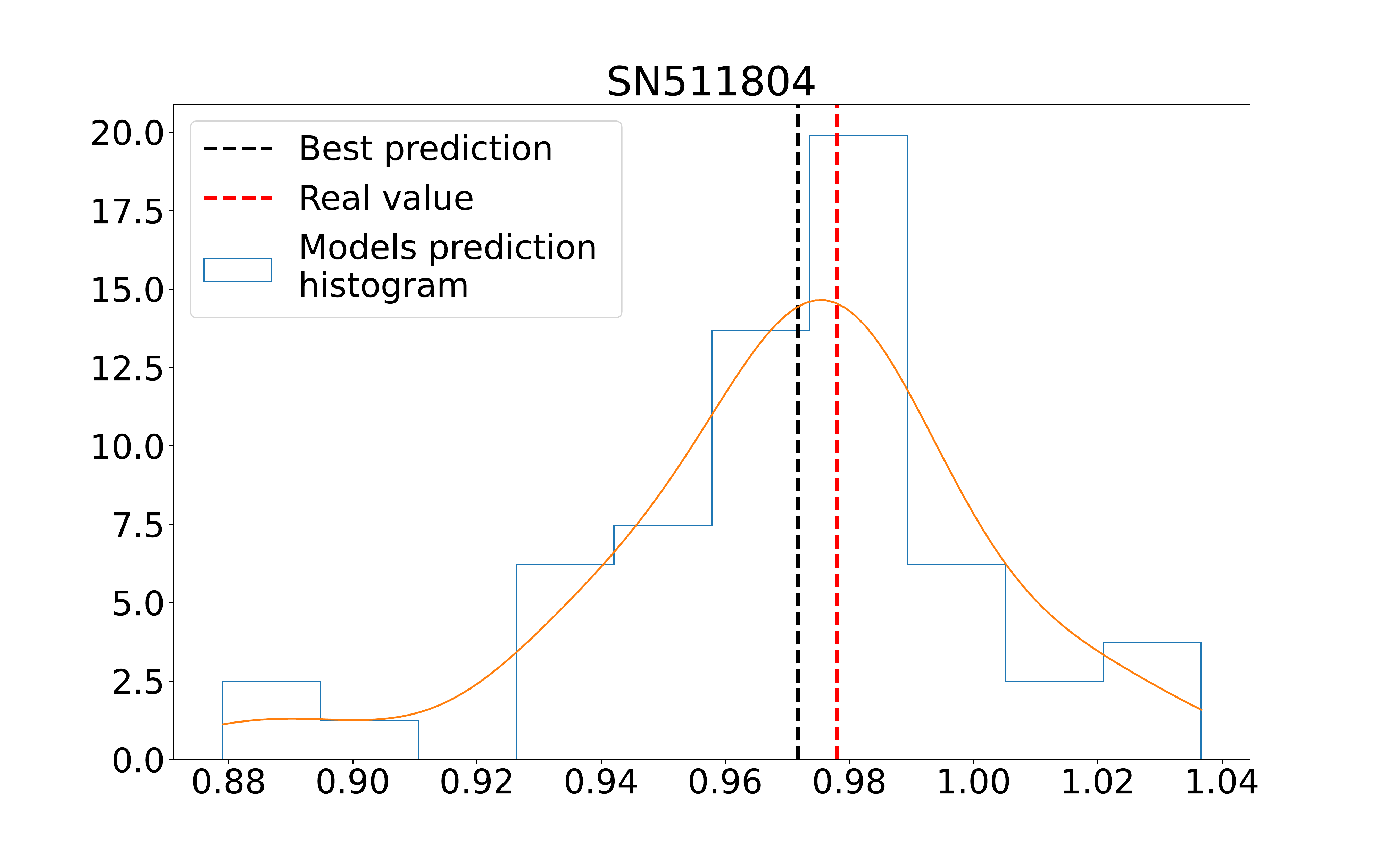}
    \includegraphics[width=\columnwidth]{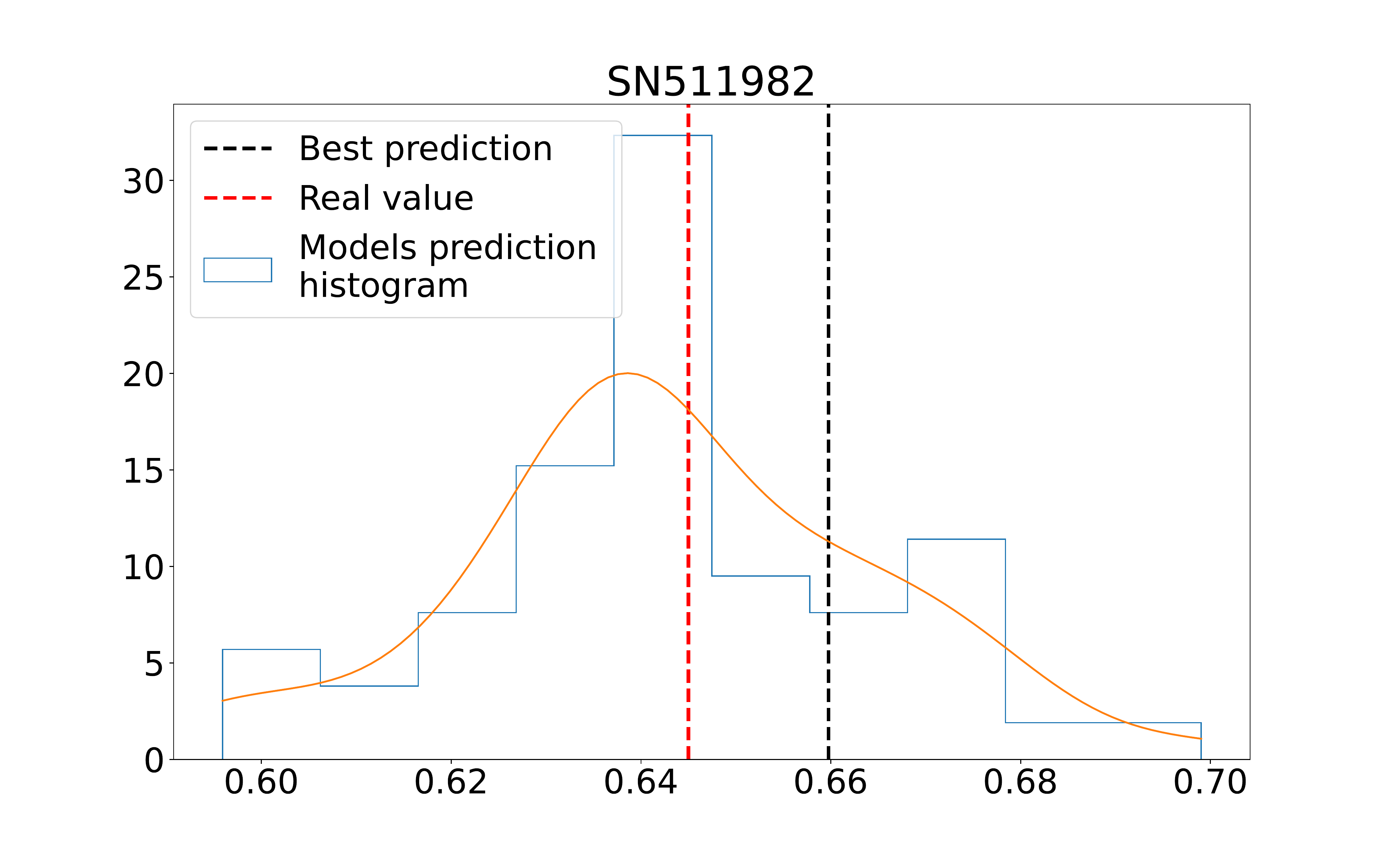}
    \caption{Probability density for SN510924 (top left), SN511436 (top right), SN511804 (bottom left) and SN511982 (bottom right).}
    \label{fig:pdfs}
\end{figure*}

From these values, we used the Scikit \citep{2020SciPy-NMeth} \textit{gaussian\_kde} method to estimate the probability density function. The graphical results are illustrated in \autoref{fig:pdfs}.

\begin{figure*}
    \centering
    \includegraphics[width=\columnwidth]{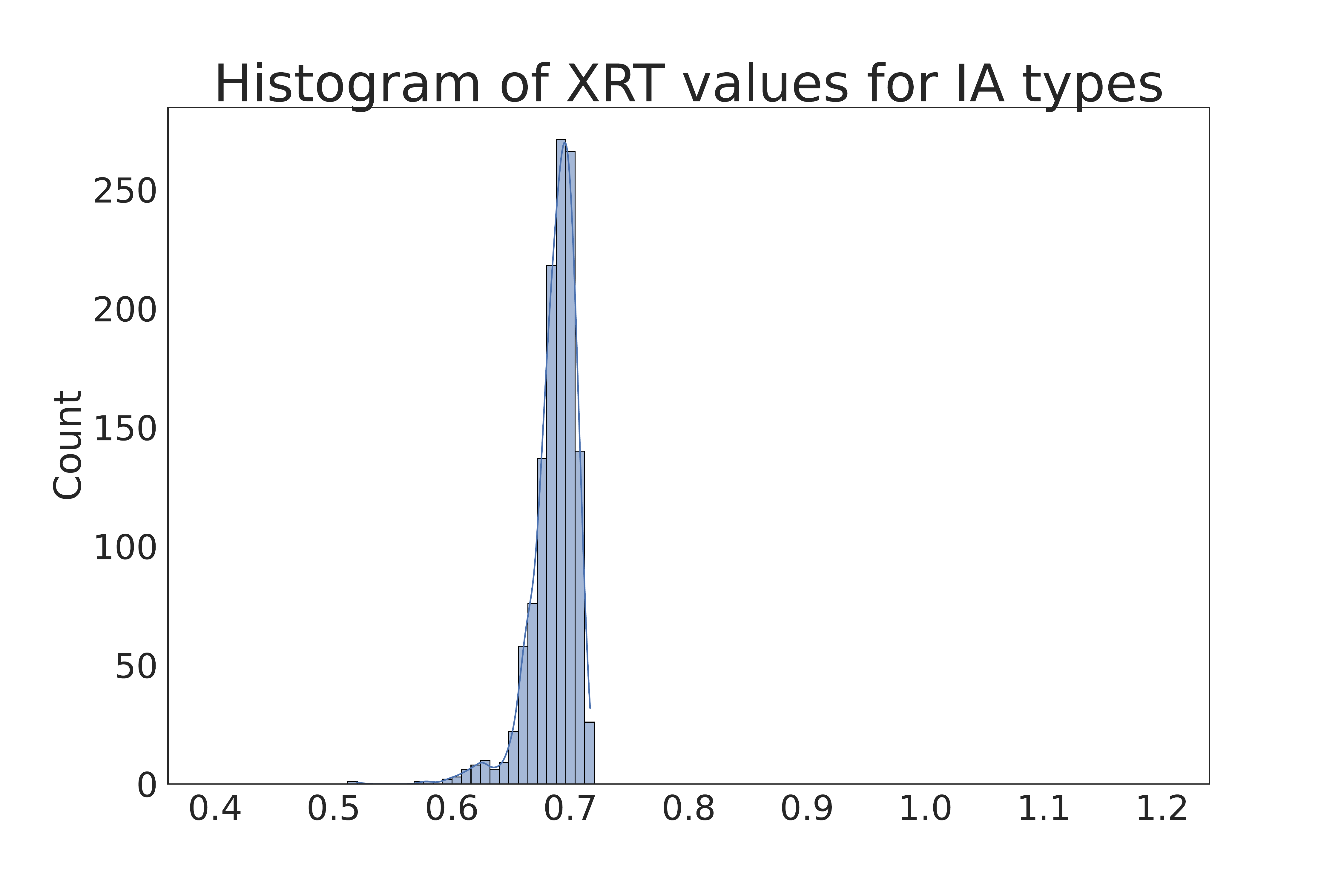}
    \includegraphics[width=\columnwidth]{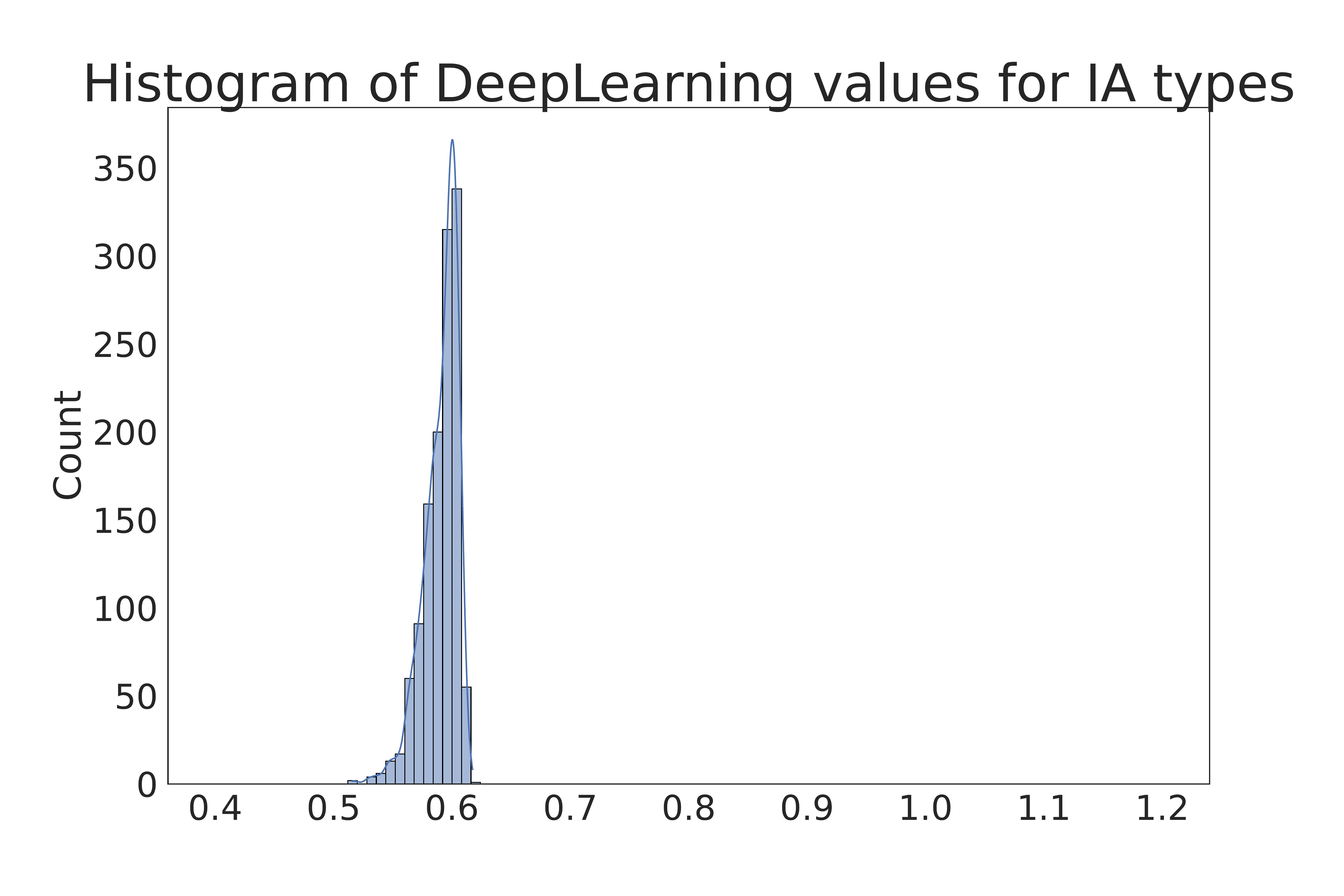}\\
    \includegraphics[width=\columnwidth]{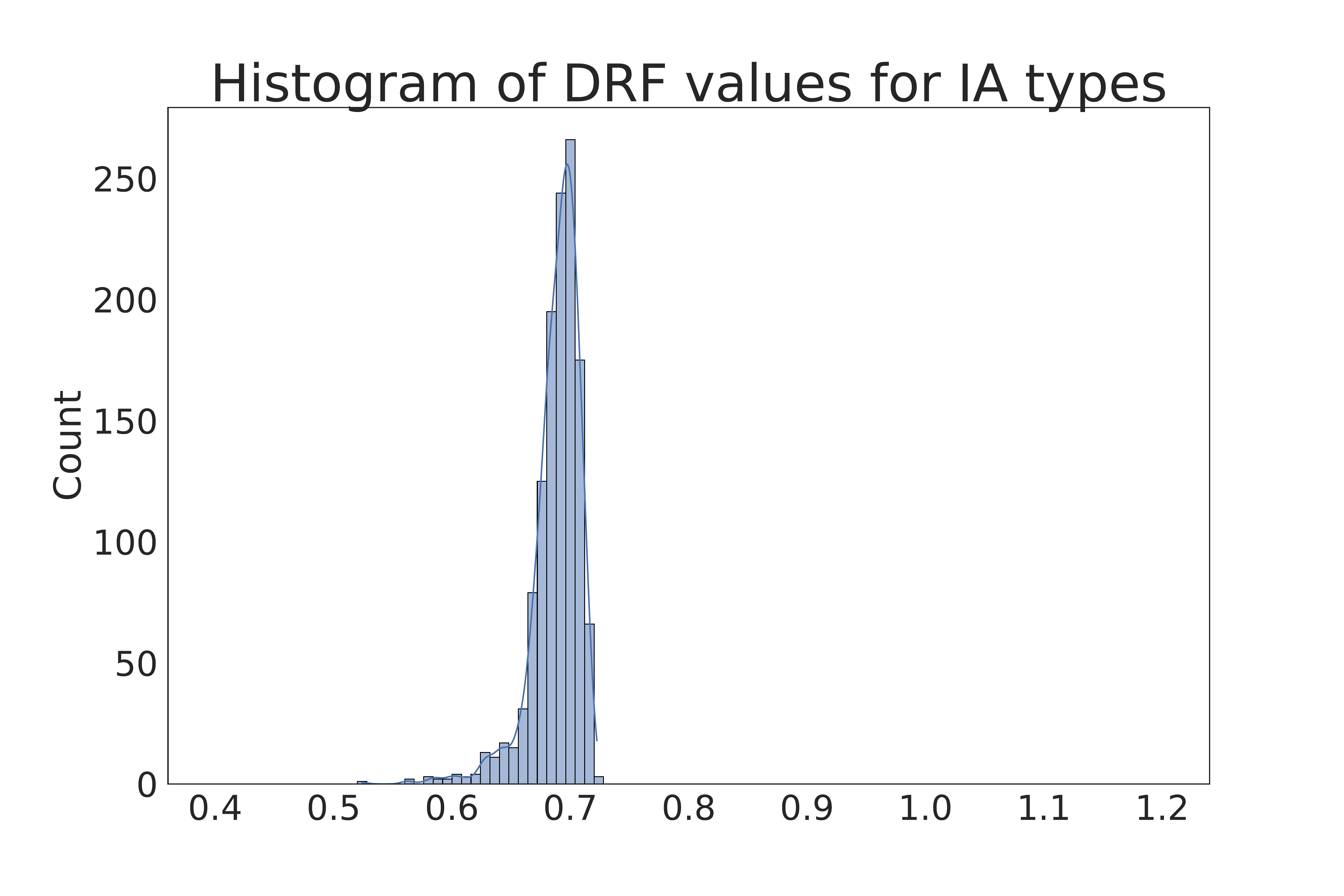}
    \includegraphics[width=\columnwidth]{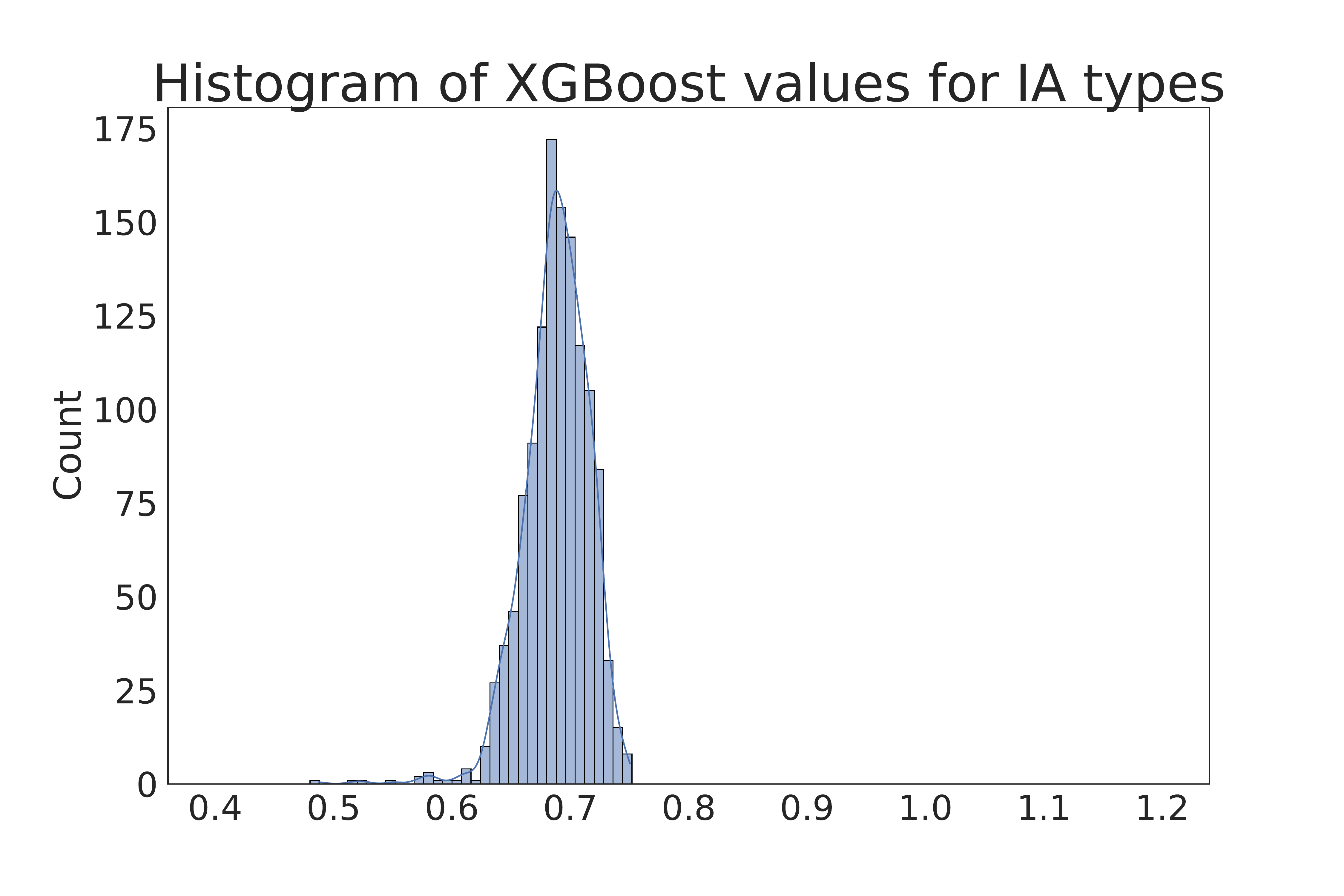}\\
    \includegraphics[width=\columnwidth]{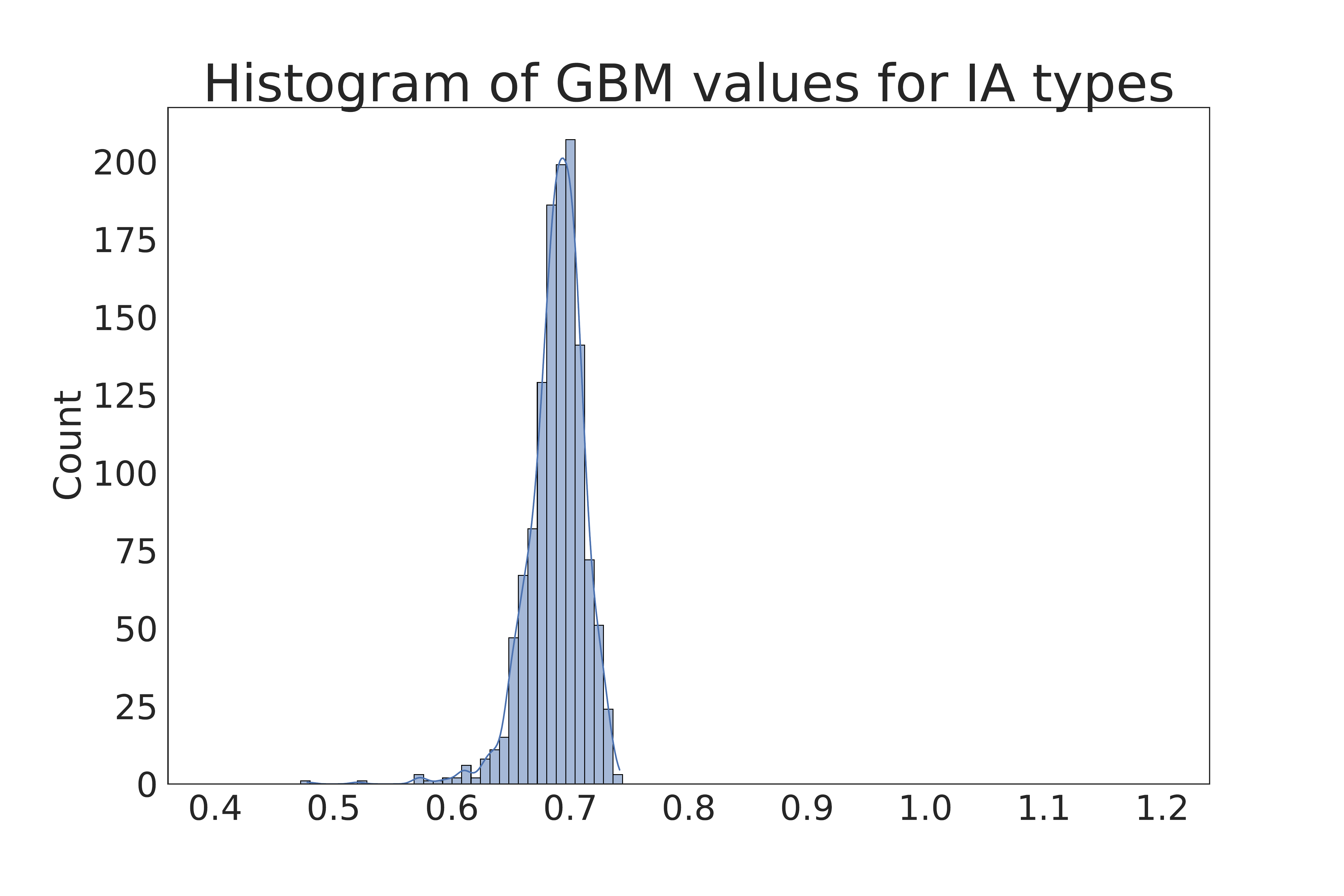}
    \includegraphics[width=\columnwidth]{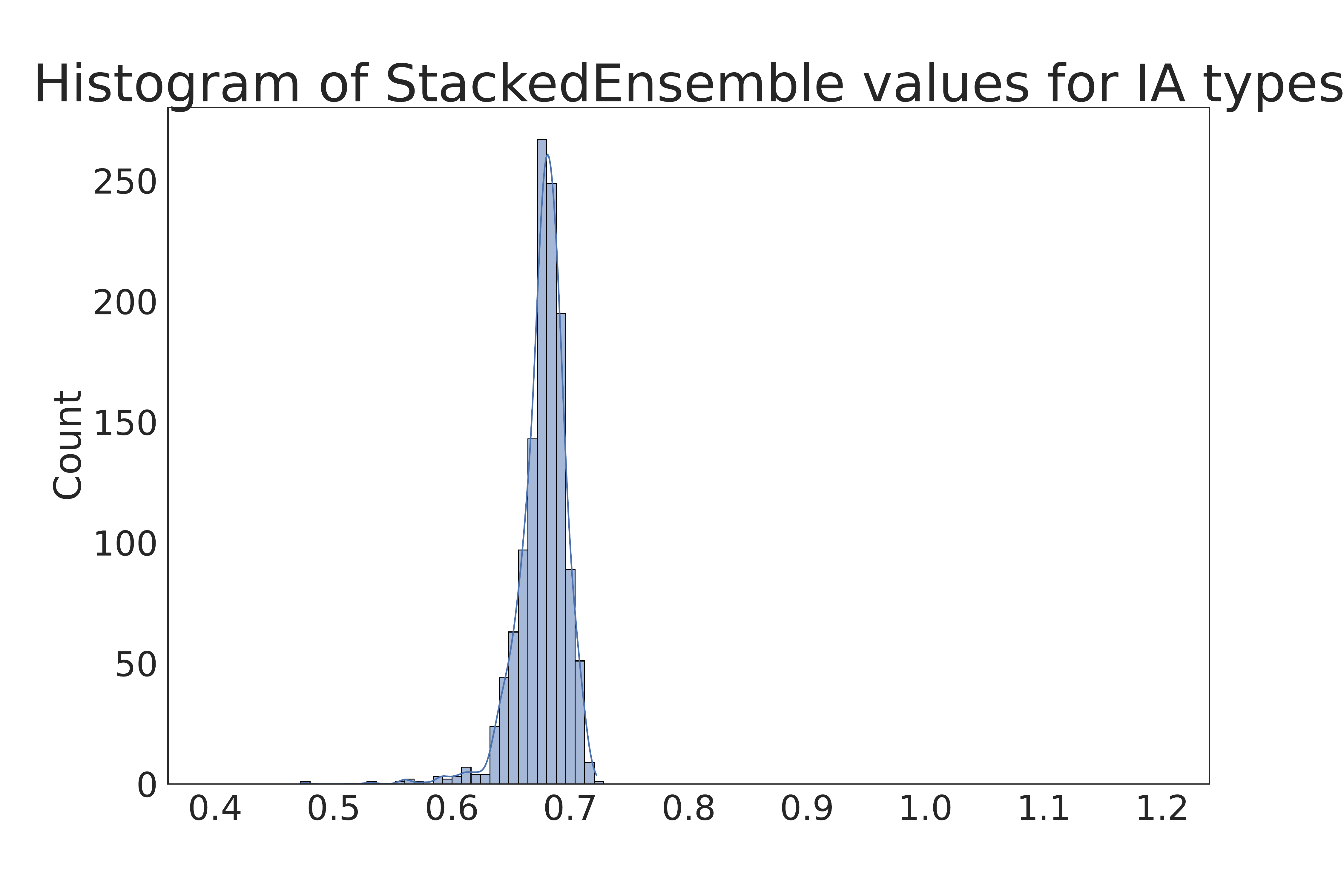}\\
    \includegraphics[width=\columnwidth]{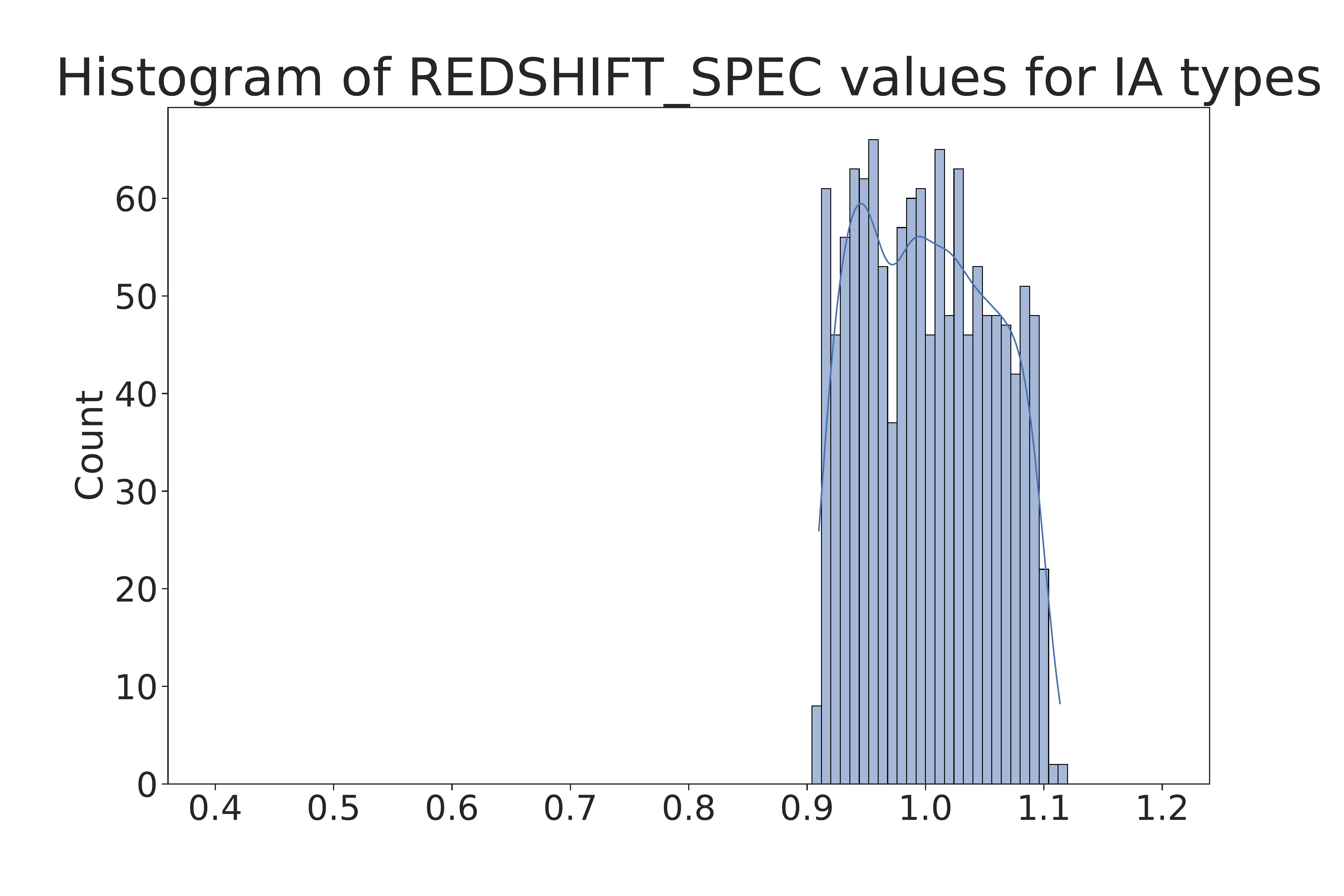}\\
    \caption{Histogram of redshift models trained with lower z predictions values for the best model from each family (Top 6 figures), and histograms with spectroscopic redshift (last figure).}
    \label{fig:lowerzvalues}
\end{figure*}

Finally, to assess the predictors in a realistic scenario, we trained a model with samples containing only the lower redshift half values in order to evaluate how the model would perform in situations where train data does not contain higher z values. Therefore, we evaluate if the model could still achieve decent results by analysing \autoref{fig:lowerzvalues}.

\subsection{Light Curve Fitting}

The template fitting\footnote{Note that \emph{Template Fitting}  refers to methods which use templates of Type Ia light curves to fit other light curves, while \emph{Template Matching} refers to ones which use templates of different types to classify any light curve.} approach consists in reconstruct the light curve of a type Ia supernova, based on templates constructed from previous data sets of SNeIa confirmed by spectroscopic surveys. We fit the light curves computing the flux of the SALT2 model by using the implementation provided by the \textsc{SNCosmo}\footnote{\url{sncosmo.readthedocs.io}} library (\url{sncosmo.readthedocs.io}), and sample 1000 points using the multinest method provided by the \textsc{PyMultinest}\footnote{\url{github.com/JohannesBuchner/PyMultiNest}} python library.

In SALT2, the rest-frame specific flux at wavelength $\lambda$ and phase (time) $p$ is
modeled by ($p = 0$ at $B$-band maximum) 
\begin{equation}
\phi(p,\lambda; x_0,x_1,c)=x_0[M_0(p,\lambda)+x_1M_1(p,\lambda)]
\exp [c \, CL(\lambda)],
\label{eq:salt2restflux}
\end{equation}
where $M_0(t,\lambda)$ is the average spectral sequence (using past supernova data, \citep{Guy:2007dv}), and $M_1(t,\lambda)$ is a higher order component related to the supernova variability. Finally, $CL(\lambda)$ is the average color correction.

The
observer-frame
flux in passband $Y$ is calculated as
\begin{equation}
F^Y(p(1+z))=(1+z)\int \phi(p,\lambda')T^Y(\lambda' (1+z)) d\lambda',
\label{eq:salt2obsflux}
\end{equation}
where $T^Y(\lambda)$ is the transmission curve of the observer-frame
filter $Y$, and $z$ is the redshift. The free parameters to be fitted are $x_0$, $x_1$ and $c$ which are the SED sequence normalization, the stretch and the color parameters respectively, and also $t_0$ (the time of B band maximum), and the supernova redshift $z$. For the light curve parameters ($x_0$, $x_1$, $c$, $t_0$) we use the priors given in \autoref{tab:salt2par}, while for the redshift, we considered three different priors, shown in \autoref{tab:salt2z}: uniform,
Gaussian with mean and variance from the host galaxy spectra and Gaussian with mean and variance from SNIa spectra.

\renewcommand{\arraystretch}{1.5}
\begin{table}
\centering
\begin{tabular}{cc}
\hline
 Parameter & Prior\\
\hline
  $t_0$ & $\mathcal{U}(-60,100)$\\
  $x_0$ & $\mathcal{U}(-10^{-3},10^{-3})$\\
  $x_1$ & $\mathcal{U}(-3,3)$\\
  $c$ & $\mathcal{U}(-0.5,0.5)$\\
\hline
 \end{tabular}
 \caption{Uniform prior ranges on the SALT2 model parameters.} \label{tab:salt2par}
 \vspace{6pt}
\end{table}

\renewcommand{\arraystretch}{1.5}
\begin{table}
\centering
\begin{tabular}{cc}
\hline
 Redshift Priors\\
\hline
  Uniform & $\mathcal{U}(0.01,1.5)$\\
  Host Galaxy $z_{phot}$ & $\mathcal{N}(z_{host}, \sigma_{host})$\\
  SNIa $z_{phot}$ & $\mathcal{N}(z_{sn}, \sigma_{sn})$\\
\hline
 \end{tabular}
 \caption{Uniform prior ranges on the SALT2 model parameters.} \label{tab:salt2z}
 \vspace{6pt}
\end{table}

\section{Results}\label{sec:res}
In this section we compare the results of stacked ensemble H2O model in relation to standard linear regression models and one of the most spread and used models nowadays, the XGBoost \citep{Chen:2016:XST:2939672.2939785}, concluding which one is the best.

In addition, we evaluated the robustness of our model by comparing scores trained with data sets of different sizes. It is a simple empirical way to illustrate the performance of the model for future uses, since in regression problems, the robustness assessment cannot be done as in classification models.

In contrast to the direct evaluation mentioned in \autoref{subsec:stdmetrics}, the objective of the robustness assessment in \autoref{subsubsec:medianabsolutedeviation} and \autoref{subsec:robustness} is to provide a magnitude order of the error considering the size of the data set used to train.

Finally a boxplot with some standard metrics provided by H2O for the 2 data arrangements we did is presented in \autoref{app:leaderboards}. Another point is that H2O's own internal methods perform better than classical regression algorithms, for example, H2O's XGBoost itself is better than the standard XGBoost (due to the hyperparameter tuning).

\subsection{Metrics}\label{subsec:metrics}

We first compare the results using root mean squared error (RMSE) and mean absolute error (MAE) metrics. After that, we use additional metrics typically used in the literature to assess how well can our prediction be used considering others recent methodologies of redshift prediction. Finally, we do a brief stacked ensemble robustness (\citep{bhagoji2017enhancing}) assessment by evaluating its performance when only trained with a few data.

To assess the performance of the redshift predictions we used the metrics bellow:

\begin{enumerate}
    \item RMSE
    \item MAE
    \item Prediction Bias $ \langle \Delta z \rangle $
    \item Median Absolute Deviation ($\sigma_{\rm MAD}$)
    \item $\eta$ outliers percentage
\end{enumerate}

Since we confirmed that stacked ensemble performs better than other models, in this subsection, we decided to do the same validation steps from \citep{Pasquet_2018} and analyze how the Auto ML model performs on smaller data set batch sizes.

The division consisted on sampling random batches of different sizes from 500 to 2500 objects for training data, and the test data - which gave the results of the metrics. It is important to mention that the data set used to generate those metrics consists only on type Ia supernovae.

The results for prediction bias ($ < \Delta z > $), Median Absolute Deviation ($\sigma$ MAD) and outliers percentage ($\eta$) for each batch are compiled in \autoref{tab:sigma_mad}.

\subsubsection{RMSE - Root Mean Square Error}\label{subsubsec:RMSE}

    Root Mean Square Error is a standard metric used to validate ML models as a whole. Its physical dimension remains the same as the target variable (redshift) and it is represented by the square root of the sum of the quadratic differences between predicted values and observed values
    
    \begin{equation}
        \mathrm{RMSE} =\sqrt{\frac{1}{N}\sum_{i=1}^N (\hat y_i - y_i)^2} \label{eq:rmse}
    \end{equation}
    
    Normally, RMSE penalizes bigger variances and outliers once it sums over the quadratic differences.

\subsubsection{MAE - Mean Absolute Error}\label{subsubsec:MAE}

    Mean Absolute Error is another standard metric used to validate ML models. Its physical dimension is also the same as the target variable. It is defined by the absolute errors between paired observations expressing the same phenomenon.
    
    \begin{equation}
        \mathrm{MAE} = \frac{1}{N}\sum_{i=1}^{N} |\hat y_{i} - y_{i} |  \label{eq:mae}
    \end{equation}
    
    Though it is very similar to RMSE, MAE does not penalize more outliers. Therefore, comparing RMSE and MAE values is a good indicative of presence of outliers and discrepancies. \footnote{For better comprehension of RMSE and MAE difference: \url{https://medium.com/human-in-a-machine-world/mae-and-rmse-which-metric-is-better-e60ac3bde13d}}

\subsubsection{Prediction Bias}\label{subsubsec:predictionbias}

    Prediction bias is an useful concept commonly used in the literature, defined as the mean of the residuals ($\Delta z$) from photometric and spectroscopic redshift.
    
    \begin{equation}
         \Delta z = \frac{z_{phot} - z_{spec}}{1 + z_{spec}}
    \end{equation}\label{eq:deltaz}
    
    \begin{equation}
         \langle \Delta z \rangle =\frac{1 }{N}\sum_{i=1}^{N} \Delta z_{i}
    \end{equation}\label{eq:predbias}

\subsubsection{Median Absolute Deviation ($\sigma$ MAD)}\label{subsubsec:medianabsolutedeviation}

    MAD deviation \autoref{eq:mad} consists on the absolute value of the difference between the residual value ($\Delta z$) and the median of residual values. The original proposal of this metric was based on \citep{2006}.
    
    We can interpret $\sigma_{\rm MAD}$ \autoref{eq:sigmamad} as a normalization based on parameters and the median of the values.
    
    \begin{equation}
         {\rm MAD} = {\rm Median}( |\Delta z - {\rm Median} (\Delta z)|) \label{eq:mad}
    \end{equation}
    
    \begin{equation}
         \sigma_{\rm MAD} = 1.4826 \times {\rm MAD} \label{eq:sigmamad}
    \end{equation} 
    
    It is important to mention that this metric and $\eta$ outliers percentage were developed inside a context of Convolutional Neural Networks evaluation. So, despite we use this metric for the same evaluation, our context and input data are completely different. In 
    \citep{Pasquet_2018} they have used images as input, while we use points from photometric simulation.
\subsubsection{$\eta$ outliers percentage}\label{subsubsec:etaoutliers}

    The $\eta$ outliers percentage is a outlier detection based on how many predictions exceeded the value of 5 times the value of $\sigma MAD$.

\subsubsection{Performance on higher redshifts by standard plot $\Delta z$ $x$ $z$}
We assess how the prediction error behaves depending on the magnitude of redshifts. The plot evaluates how the models perform at higher redshifts and visually shows how errors are caused due to higher $z$ values and how the absence of photometric input data influence the results.

\begin{figure*}
    \centering
    \includegraphics[width=2.\columnwidth]{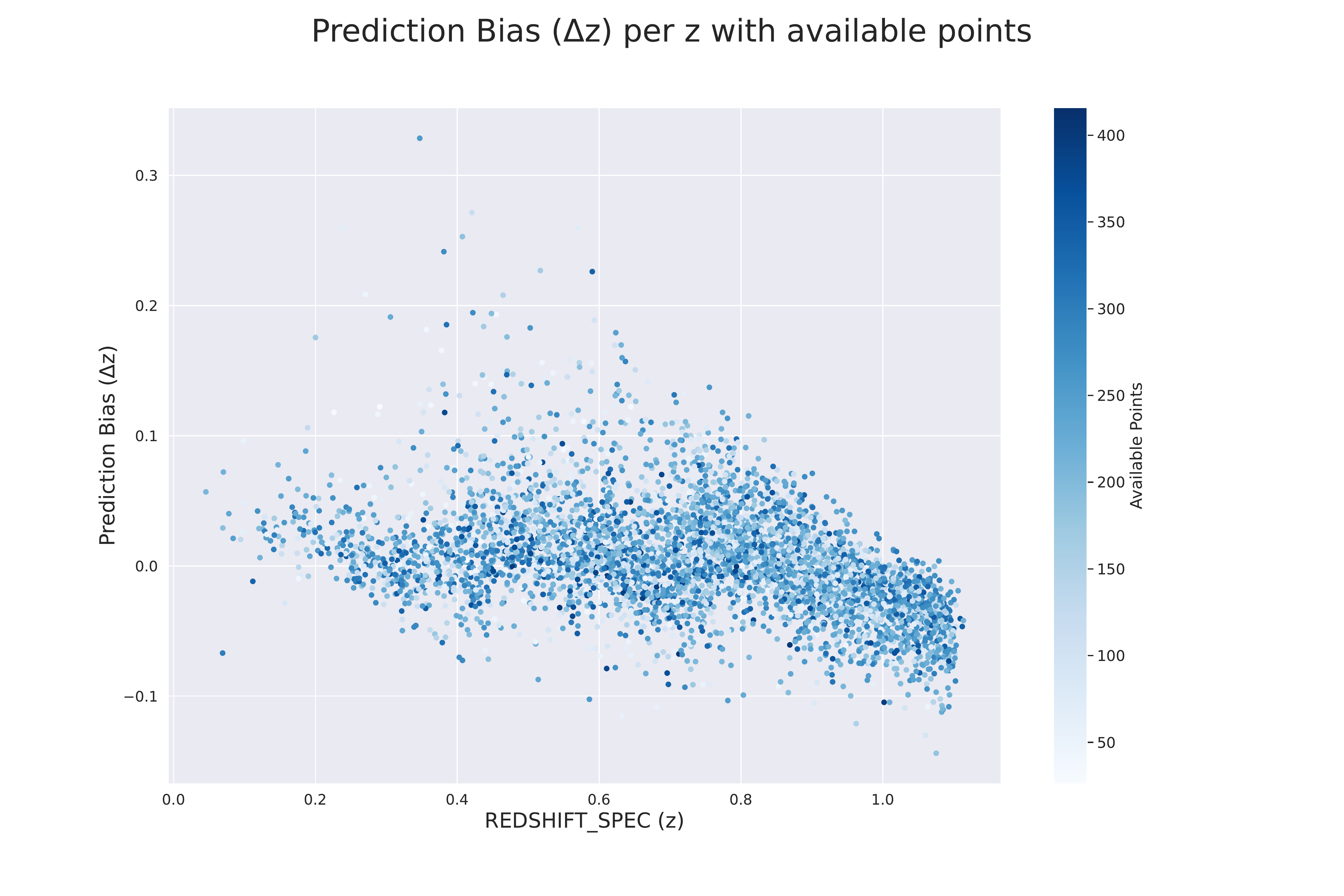}
    \caption{Prediction Bias evolution along redshift values. Scatter plot is also illustrating the amount of photometric points used for each prediction.}
    \label{fig:deltazz}
\end{figure*}

\begin{figure*}
    \centering
    \includegraphics[width=2.\columnwidth]{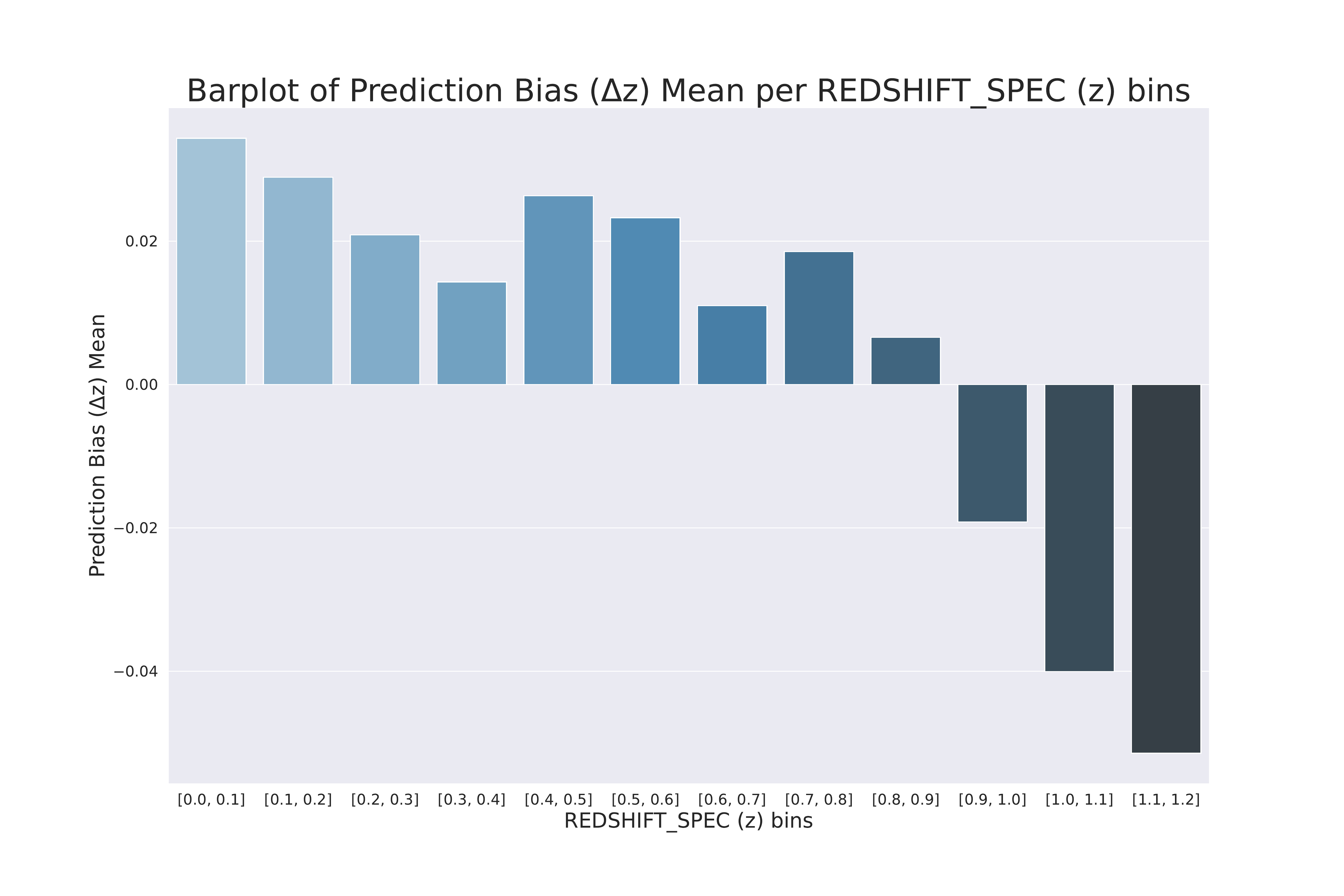}
    \caption{Prediction Bias mean evolution along redshift bins.}
    \label{fig:barplot_prediction_bias}
\end{figure*}

As this evaluation metric is entirely visual, the results are presented on \autoref{fig:deltazz} and \autoref{fig:barplot_prediction_bias} and the conclusions presented on \autoref{sec:con}.

\subsection{ML Standard models}\label{subsec:stdmetrics}

In \autoref{tab:metrics_IA} we compare the results from stacked ensemble predictions with some standard models from Machine Learning literature through RMSE and MAE metrics.

When analyzing them, we considered 2 different data sets, the first with only Ia type supernovae objects  and the second one with all objects from SNPCC. All models used techniques of cross validations  \citep{hastie2009} following the proportion of 80-20 to validate the result, then RMSE and MAE are the mean of the 5 folds.

The models are:
\begin{enumerate}
    \item Linear Regression \citep{LinearRegression}
    \item Ridge Regression \citep{Lasso}
    \item Lasso Regression \citep{RidgeRegression}
    \item Elastic Net \citep{ElasticNet}
    \item XGBoost \citep{chen2015xgboost}(best model before testings)
\end{enumerate}

\begin{table*}
\centering
\begin{tabular}{@{}lcccc@{}}
\hline
& \multicolumn{2}{c}{Type Ia} & \multicolumn{2}{c}{All Types}\\
Model & RMSE & MAE & RMSE & MAE \\
\hline
Linear Regression             & 0.27451           & 0.14155 & 0.28458 & 0.18735\\
Ridge Regression              & 0.27451           & 0.14155 & 0.28458 & 0.18735\\
Ridge Regression normalized   & 0.20095           & 0.16712 & 0.22526 & 0.19078\\
Elastic Net                & 0.22795           & 0.18927 & 0.23650 & 0.19968\\
Lasso Regression              & 0.22795           & 0.18927 & 0.23650 & 0.19968 \\
XGBoost                       & 0.16360           & 0.13202 & 0.20292 & 0.16843 \\
AutoML H2O - stacked ensemble & 0.08656           & 0.06632 & 0.13982 & 0.10429 \\
\hline
\end{tabular}
\caption{RMSE and MAE for data with all type of supernovae.} \label{tab:metrics_IA}
\end{table*}

\begin{table}
\begin{tabular}{llll}
\hline
Batch size & $\langle \Delta z \rangle$ & $\sigma_\text{MAD}$   & $\eta$    \\
\hline
500        & -0.006887 & 0.0611498 & 0.4119702 \\
1000       & 0.0019815 & 0.0493086 & 0.3182932 \\
1500       & -0.002261 & 0.0464158 & 0.2948574 \\
2000       & -0.002692 & 0.0444166 & 0.2777777 \\
2500       & 0.0025125 & 0.0349551 & 0.1867088 \\
\hline
\end{tabular}
\caption{Stacked ensemble performance. Prediction bias, $\sigma_{\rm MAD}$ and outliers percentage type IA supernovae of different sizes.} \label{tab:sigma_mad}
\end{table}

\subsection{Robustness Evaluation}\label{subsec:robustness}

We show in \autoref{tab:sigma_mad} the results for the prediction bias, $\sigma_{\rm MAD}$ and outliers percentage, as a function of the training sample size. We can see that the bias is bellow $10^{-2}$ for all tested sizes, while there is a significant decrease in $\sigma_{\rm MAD}$ and $\eta$ for larger sizes.

As H2O stacked ensemble model proved to be the best model, we will analyse its robustness by training with only few data points, splitting the original type Ia supernovae data set into subsamples with sizes from 500 to 2500.

We can see in \autoref{tab:robustness} an increase of $\sim$10\% in RMSE and in MAE when augmenting the data. So in future uses of this model it is expected that the model keeps the error inside that range for training samples considerably smaller than the photometric one, thus attesting the model robustness in terms of maintaining resilience inside that range of error scores.

As a final model characteristic, we can see that the RMSE and the MAE do not present great differences. Through this, we can conclude that the algorithm has a roughly constant performance regardless of whether the values are outliers or not.

We show in \autoref{tab:salt-scores} the same metrics defined in \autoref{subsec:metrics} applied for the SALT2 light curve parameters $x_1$ and $c$. All metrics are bellow $10^{-7}$ for $x_0$. We can see that choice for the redshift prior has little impact on the results for the others parameters.

{
\begin{table}
\begin{tabular}{lll}
\hline
Batch size & RMSE     & MAE    \\
\hline
500        & 0.09961 & 0.07868 \\
1000       & 0.09651 & 0.07648 \\
1500       & 0.09437 & 0.07414 \\
2000       & 0.09211 & 0.07147 \\
2500       & 0.09125 & 0.07056 \\
\hline
\end{tabular}
\caption{Stacked ensemble performance. RMSE and MAE data set for type IA supernovae of different sizes.} \label{tab:robustness}
\end{table}
}

{
\begin{table}
\begin{tabular}{lrrrrrr}
\hline
Prior & \multicolumn{2}{c}{Uniform} & \multicolumn{2}{c}{SNIa} & \multicolumn{2}{c}{Galactic} \\
Parameter &      $x_1$ &     $c$ &    $x_1$ &     $c$ &       $x_1$ &     $c$ \\
\hline
Prediction bias      &    0.02 &  0.04 &  0.04 &  0.03 &     0.04 &  0.03 \\
MAD       &    0.45 &  0.07 &  0.46 &  0.06 &     0.46 &  0.05 \\
$\sigma_{\rm MAD}$ &    0.67 &  0.11 &  0.68 &  0.09 &     0.68 &  0.08 \\
$\eta$  &    0.92 &  0.61 &  0.92 &  0.57 &     0.93 &  0.53 \\
\hline
\end{tabular}
\caption{Scores for the SALT2 parameters. All scores are less than $10^{-7}$ for $x_0$.} \label{tab:salt-scores}
\end{table}
}

\section{Conclusions}\label{sec:con}

In this paper we have worked on the problem of supernova redshift prediction using photometric light curve data. The raw data was a simulated open source data set from Supernova Photometric Classification Challenge (SNPCC). We assessed how different machine learning models performed on that problem, specially the Auto ML models from H2O framework and their stacked ensemble methods. We compared the results from traditional models and from more advanced algorithms. They were implemented on XGBoost library and inside H2O module, verifying that the stacked ensemble methods performed best. This family of methods is an ensemble model of strong learners, also known as super learner, which gathers other models like Gradient Boosting, XGBoost, DeepLearning based algorithms and Distributed Random Forests; all of them with optimized hyperparameters thanks to the grid search optimization.

We compared the results using standard metrics like RMSE and MAE, aiming to assess the results in a way that could provide information of the model's behavior considering outliers and gross errors. Then we concluded that stacked ensemble method was by far the best in every category and it was equally robust when dealing with outliers.

Thereupon, we have explored the metrics $ \sigma$ MAD and $\eta$ outliers in order to compare and evaluate how the Auto ML model performed on other metrics used in cosmology. We also alternated the train batches sizes so we could evaluate its robustness in a sense of data quantity. Despite not having performed better than the previous results in the literature that presented such metrics, these results they serve as a baseline for future AutoML uses, besides, they used different types of data sets, using galaxies images directly on a neural network, rather than photometric points of the light curve. Thereby, it is not recommendable to compare meticulously different techniques using different amount of data sets from different natures (galaxies images vs spectroscopy light curve points).

\autoref{fig:lowerzvalues} shows how, in terms of error and bias, the models could not predict higher redshift values if they were not trained with data representing those higher $z$ cases.

The \autoref{fig:deltazz} shows no notable correlation between $\Delta z$, $z$ and the available photometric data points. We also see that the bias tends to decrease, reaching some negative values once $z$ gets higher.

\begin{figure*}
    \centering
    \includegraphics[width=2.\columnwidth]{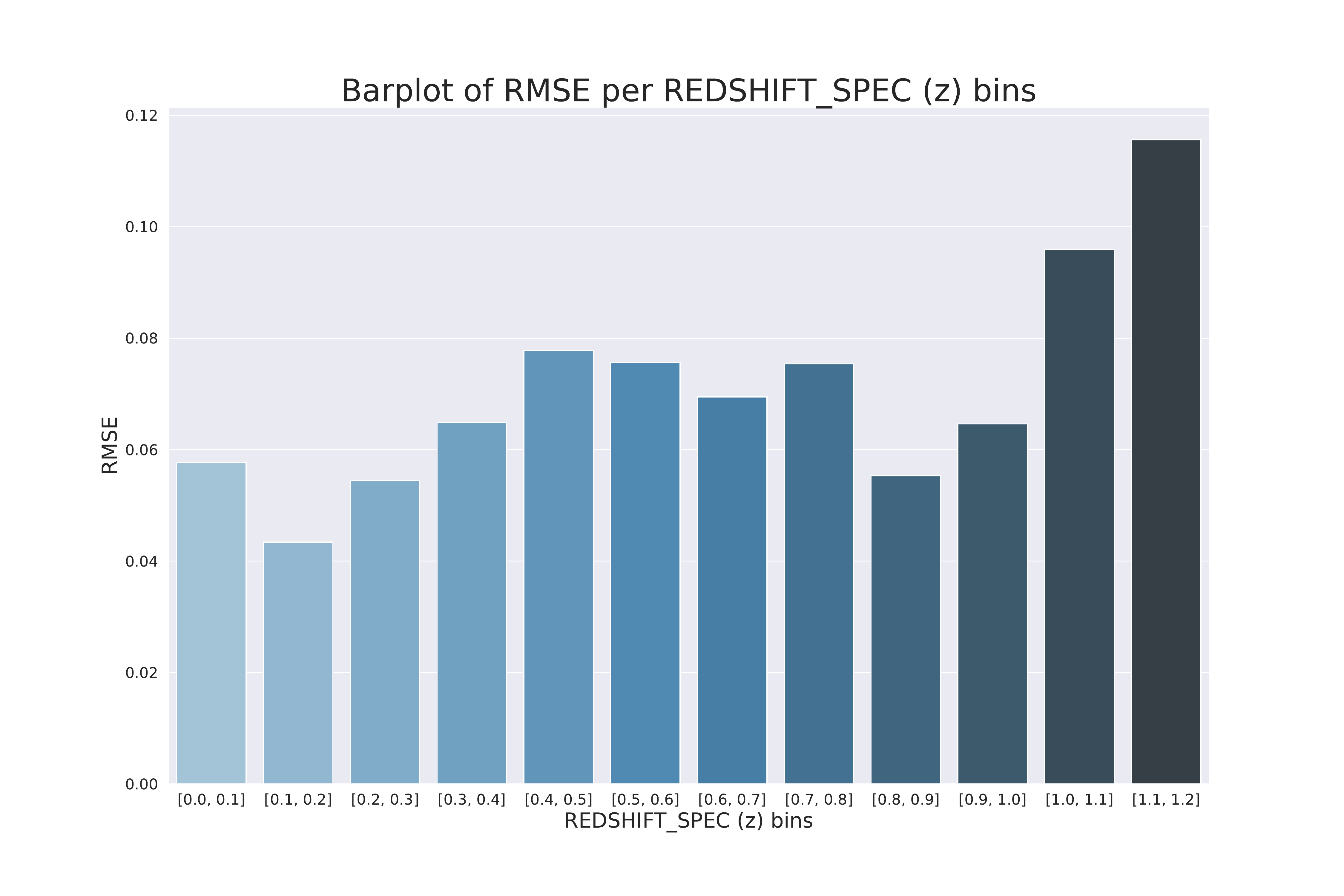}
    \caption{Barplot of RMSE values from predictions per redshift bin values.}
    \label{fig:rmse_barplot}
\end{figure*}

To get a better perspective of the errors, we presented \autoref{fig:barplot_prediction_bias} and \autoref{fig:rmse_barplot} as two simplistic barplots based on the means of the $\Delta z$ and RMSE respectively, per bins of 0.1. From samples whose redshifts stay between 0.0 and 1.0 we could no conclude that there is a simple increase ou decrease in the prediction bias or RMSE. While redshifts above 1.0 present higher RMSE and higher $\Delta z$ absolute value than redshifts from a range 0.0 to 1.0.

From the standpoint of data quality, we can conclude that the error is unaffected by the amount of photometric data points used for prediction (\autoref{fig:deltazz} color scale), but it is influenced by the range of $z$ values seen during training (shown in figure \ref{fig:lowerzvalues}). Finally, we can conclude that those techniques should not be used to predict a range of $z$ values that were not provided during the training, yet, they are resilient to predictions with short photometric data and they present an augmentation on prediction bias and RMSE for $z$ values over than 1.

The next evaluation step consisted on assessing the model robustness by alternating the train batches sizes and analyzing the MAE and RMSE, providing machine learning oriented robustness assessment and a notion of capacity of performance for future uses.

As we have worked with an ensemble method composed by strong learners, it was possible to generate PDFs to input on a SALT2 model. Each result of the strong learners that composed the final stacked ensemble method created a Gaussian KDE interpolation, enabling to get a redshift probability density function of the result, as we can see in \autoref{fig:pdfs}. Based on the similar scores \autoref{tab:salt-scores} we conclude that we can achieve essentially the same performance using the PDFs obtained in this work instead of the host galaxy information. This is particularly interesting for the distant supernovae for which the host may not be identifiable.

\section*{Acknowledgements}

MVS acknowledges PRONEX/CNPq/FAPESQ-PB (grant no. 165/2018). RRRR acknowledges CNPq (grant no. 309868/2021-1).

\section*{Data Availability}

\begin{enumerate}
    \item Code source at: \url{https://github.com/FelipeMFO/photometric-redshift-prediction}
    \item Data generated (models, pickles and csvs) at:  \url{https://drive.google.com/drive/folders/1wDXYUQ8cPdmdSigeg2kieripRyj9lM0D?usp=share_link}
\end{enumerate}







\appendix
\section{Leaderboards}\label{app:leaderboards}

\begin{figure*}
\includegraphics[width=1.75\columnwidth]{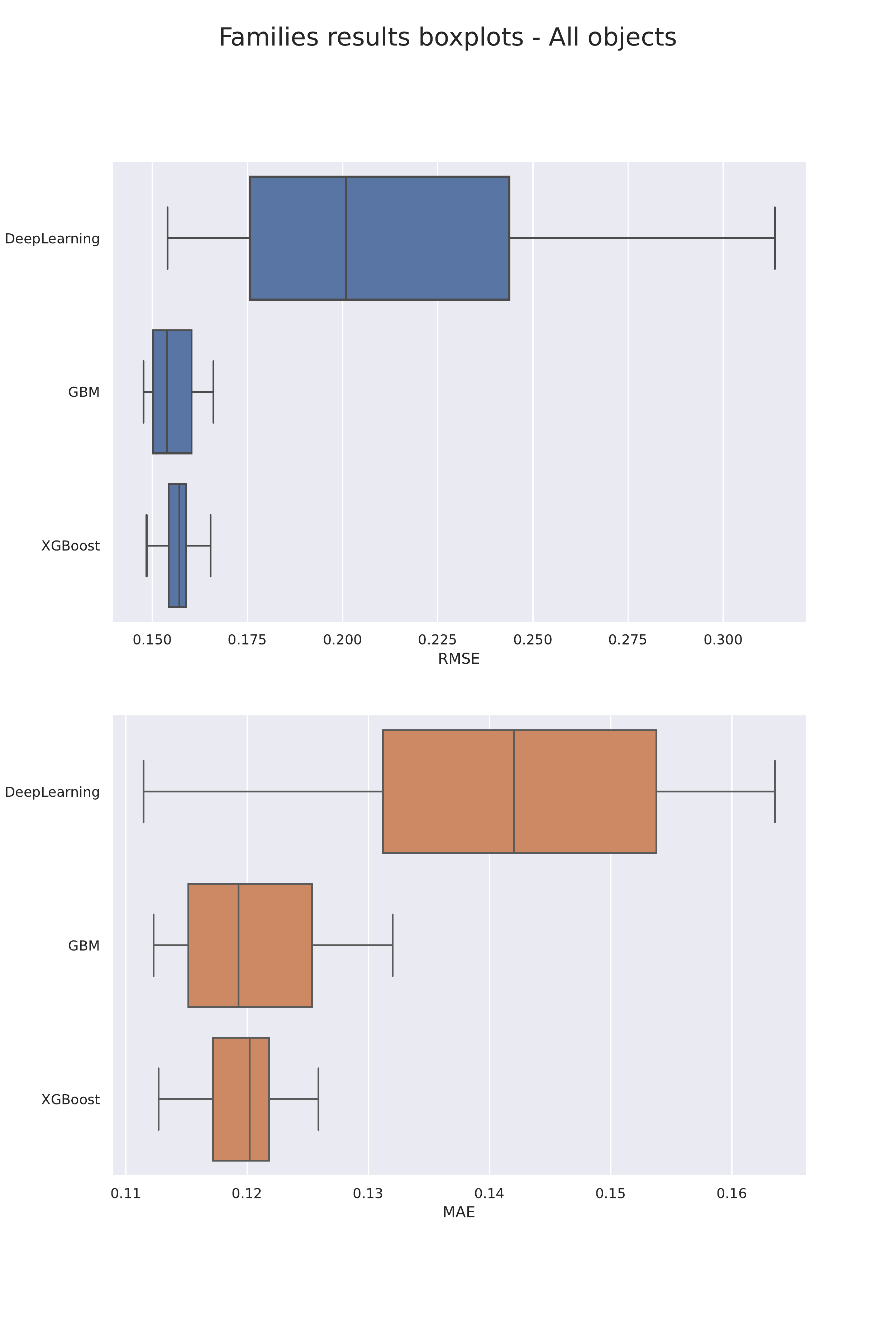}
\caption{Models` families trained on AutoML. Result of 47 models for all objects (21 XGBoost, 15 GBM and 11 DeepLearning, families with 2 or less models were ignored)}
\label{fig:amlboard50}
\end{figure*}

\begin{figure*}
\includegraphics[width=1.75\columnwidth]{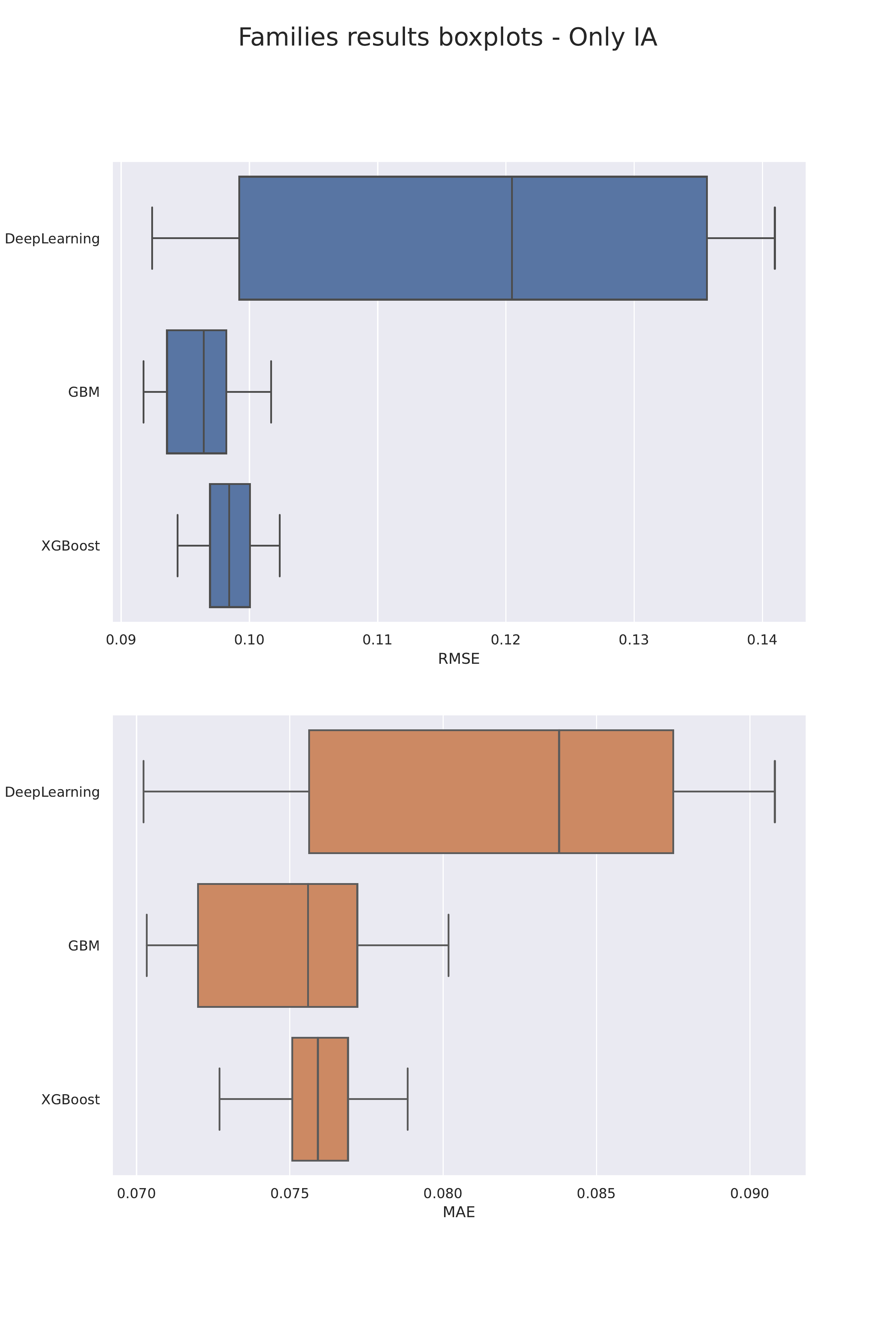}
\caption{Models` families trained on AutoML. Result of 47 models for only IA objects (21 XGBoost, 15 GBM and 11 DeepLearning, families with 2 or less models were ignored)}
\label{fig:amlboard50IA}
\end{figure*}


\bsp	
\label{lastpage}
\end{document}